\documentclass{article}

\usepackage{PRIMEarxiv}

\usepackage[utf8]{inputenc} % allow utf-8 input
\usepackage[T1]{fontenc}    % use 8-bit T1 fonts
\usepackage{hyperref}       % hyperlinks
\usepackage{url}            % simple URL typesetting
\usepackage{booktabs}       % professional-quality tables
\usepackage{amsfonts}       % blackboard math symbols
\usepackage{nicefrac}       % compact symbols for 1/2, etc.
\usepackage{microtype}      % microtypography
\usepackage{fancyhdr}       % header
\usepackage{graphicx}       % graphics
\usepackage{pdflscape}
\usepackage[table,xcdraw]{xcolor}
\usepackage{caption}
\usepackage{colortbl}
\usepackage{longtable}
\usepackage{array}
\graphicspath{{media/}}     % organize your images and other figures under media/ folder
\usepackage{longtable,booktabs,array}
\usepackage{calc} % for calculating minipage widths
\usepackage{etoolbox}
\makeatletter
\patchcmd\longtable{\par}{\if@noskipsec\mbox{}\fi\par}{}{}
\makeatother
\IfFileExists{footnotehyper.sty}{\usepackage{footnotehyper}}{\usepackage{footnote}}
\makesavenoteenv{longtable}

\newlength{\cslhangindent}
\newlength{\csllabelwidth}
\newlength{\cslentryspacingunit} % times entry-spacing
  {}%
  {\par}
\newenvironment{CSLReferences}[2] % #1 hanging-ident, #2 entry spacing
 {% don't indent paragraphs
  \setlength{\parindent}{0pt}
  \ifodd #1
  \let\oldpar\par
  \def\par{\hangindent=\cslhangindent\oldpar}
  \fi
  \setlength{\parskip}{#2\cslentryspacingunit}
 }%
 {}

\title{Dynamic Time Warping-based imputation of long gaps in human
mobility trajectories
}

\author{
  Danielle McCool, Peter Lugtig \\
  Methodology and Statistics \\
  Utrecht University\\
  Utrecht, Netherlands\\
  \texttt{d.m.mccool@uu.nl, p.lugtig@uu.nl} \\
   \And
  Barry Schouten \\
  Statistics Netherlands, Utrecht University \\
  Den Haag, Netherlands\\
  \texttt{jg.schouten@cbs.nl} \\
}

\begin{document}
\maketitle

\begin{abstract}
 Individual mobility trajectories are difficult to measure and often
  incur long periods of missingness. Aggregation of this mobility data
  without accounting for the missingness leads to erroneous results,
  underestimating travel behavior. This paper proposes Dynamic Time
  Warping-Based Multiple Imputation (DTWBMI) as a method of filling long
  gaps in human mobility trajectories in order to use the available data
  to the fullest extent. This method reduces spatiotemporal trajectories
  to time series of particular travel behavior, then selects candidates
  for multiple imputation on the basis of the dynamic time warping
  distance between the potential donor series and the series preceding
  and following the gap in the recipient series and finally imputes
  values multiple times. A simulation study designed to establish
  optimal parameters for DTWBMI provides two versions of the method.
  These two methods are applied to a real-world dataset of individual
  mobility trajectories with simulated missingness and compared against
  other methods of handling missingness. Linear interpolation
  outperforms DTWBMI and other methods when gaps are short and data are
  limited. DTWBMI outperforms other methods when gaps become longer and
  when more data are available.\end{abstract}

\hypertarget{introduction}{%
\section{Introduction}\label{introduction}}

The mobile device as a new source of data for researchers has promised
returns across a variety of fields including human geography,
transportation and mobile health (Rout et al., 2021). These gains have
yet to materialize due to a lack of generic resources available to
handle the frequent data quality issues, especially with respect to
missing data (Bähr et al., 2020; Beukenhorst et al., 2022). Studies
using standalone Global Navigation Satellite System (GNSS) tracking
devices have demonstrated the usefulness of trace data in collecting
activity data, in predicting wandering episodes in adults with dementia
and in investigating travel behavior (Furletti et al., 2013; Wojtusiak
and Nia, 2021; Zheng et al., 2008).

Researchers presupposed a natural progression from these standalone
devices to the use of respondents' own devices as an economic and
flexible alternative. This is within the technical capacity of the
smartphone, but the reality of this data collection methodology is that
it is highly prone to missingness (Harding et al., 2021; Yoo et al.,
2020). In fact the same feature that makes it compelling for researchers
-- collection of large amounts of data over time -- becomes a
significant drawback when the data are incomplete. The temporal nature
of these data make straightforward imputation of the exact missing
locations an intractable problem. In this paper, we propose and evaluate
an alternative method for imputation of longer sections of missing
location data in travel surveys. Dynamic Time Warping-based Imputation
uses aggregate statistics derived from continuously measured locations
to impute travel metrics during the gap, allowing full use of the
information that precedes and follows the gap.

Data can be missing for many reasons, some related to the physical
surroundings and thus common to all GNSS measurements, and others
related to the device, the user, or the interaction between the two
(Bähr et al., 2020; Ranasinghe and Kray, 2018). The first category
contains such mechanisms as line-of-sight problems, cold starts and
urban canyons (Karaim et al., 2018). These tend to produce missingness
that is consistent, occurring either over a limited geographic range or
within a relatively brief period of time. On the other hand, the second
category varies significantly between persons and their devices
(Beukenhorst et al., 2021). Storing the location data requires that an
app must be running on the smartphone, the smartphone must be on, and it
must have the necessary permissions to record the data (Keusch et al.,
2022). Both Android and iOS operating systems, in an effort to increase
battery life on their devices, will often shut down apps
indiscriminately, including those applications recording data for
research purposes. This is a primary cause of missing data collected
from smartphones (González-Pérez et al., 2022). In addition to this
largest hurdle, apps must also contend with the user closing the app,
turning off the phone or disabling the location services on the device.
This can occur unintentionally, but may occasionally be due to privacy
concerns (Kreuter et al., 2020) or concerns about battery life. Lastly,
the battery may simply become drained, ceasing all location collection
until the device is turned on again.

Developing strategies to handle the missing data is as necessary as it
is challenging (Moffat et al., 2007). Studies have demonstrated that the
patterns of missing data are more often informative than not (Bähr et
al., 2020; Mennis et al., 2018). Consequently, deletion of those persons
or days with missing data can produce biased results in subsequent
analyses (Hawthorne and Elliott, 2005; Honaker and King, 2010). Neither
is it appropriate to interpolate across gaps of any appreciable size,
leading to a considerable rate of underestimation as gaps lengthen to
include whole trips (McCool et al., 2022; Phan et al., 2018a) and up to
a 10-fold increase in error variance computed on the metrics of interest
(Barnett and Onnela, 2020). It is possible to aggregate the data to days
or weeks and then to apply more traditional forms of missing data
handling. This is the method most commonly employed by accelerometer
studies (Stephens et al., 2018). However, this loss of granularity
restricts the level of analysis to the aggregate level. More
sophisticated methodologies are necessary to simultaneously maintain the
useful characteristics of the data as well as allow for unbiased
handling of missingness (Onnela, 2021).

These sophisticated methodologies do exist in the literature, but so far
none have demonstrated the capacity to solve long gaps in the data
without problems. For example, k-Nearest Neighbor (kNN) methodologies
are often employed to impute gaps in this way, but tend to
insufficiently consider the time aspect of univariate time series (Sun
et al., 2017). Time-Delayed Deep Neural Networks have recently sown some
promise for the imputation of long gaps, but currently consider only a
single gap in data supported by related variables (Park et al., 2022).
Efforts to reconstruct full sets of potential paths with adequate
coverage have been considered recently which may prove interesting, but
may also lead to huge sets of possibilities when considering long gaps
in human mobility (Parrella et al., 2021). Random walks may be used to
generate sequences of behavior that may be used for gap-filling across
long gaps, but current methodologies are still in development (Dekker et
al., 2022).

Geographic or time constraints can help to identify some trajectories
that are more likely, and to narrow the number of possible tracks.
However, this requires additional data that limits their applicability
in the general situation. For example, we may make use of the repetitive
nature of human behavior and the fact that two weeks tend to contain a
person's primary activity spaces (Stanley et al., 2018; Zhu et al.,
2022) if we extend data collection over a long enough temporal period to
achieve coverage over the missing data periods (Chen et al., 2019; Dhont
et al., 2021). Or, if a geographical constraint can be applied, routes
that make use of public transportation or common road structures may be
shared with a high frequency. Users may then have overlapping traces
allowed for individuals to complete each others' records with a high
degree of precision.

External data has also been shown to provide additional benefits in the
handling of missing data. Map-matching may provide a method for
realistically imputing travel patterns provided that the gap is short
enough that it can account for both the start and stop locations
(Jagadeesh and Srikanthan, 2017; Knapen et al., 2018; Tanaka et al.,
2021) or if the missed locations can be provided after the fact by users
or by data donation (Boeschoten et al., 2020; Hollingshead et al., 2021;
Keusch and Conrad, 2022; Silber et al., 2021). It is possible that data
on mode of transportation is useful in filling long gaps where the
travel behavior depends strongly on the mode, but asking for this
information increases the burden on respondents and is prone to error.
Importantly, each of these methodologies requires specific additional
data or constraints that are unlikely to be compatible with the low
density and short observation periods common for travel studies.

Often there may be no helpful relationships within the data to allow for
the prediction of missing data beyond what patterns exist within and
between individuals in the recorded data. Dynamic Time Warping is a
method used to find similar areas within two time series, which we
propose as a selection mechanism for multiple imputation candidates on
the basis of the location data itself. This paper 1) presents Dynamic
Time Warping-Based Imputation, a new generically applicable methodology
for use in this scenario, and 2) evaluates its performance against other
applicable methodologies.

The next section will describe and outline the integral elements of this
methodology, including extraction of metrics of interest as time series,
the calculation of dynamic time warping distances between these time
series, followed by multiple imputation using this distance measure to
select candidates for donation of their time series, and a description
of Dynamic Time Warping-Based Imputation. We describe the history of
this methodology and outline our adaptations of the method for use in
the imputation of geolocations over time. Section \ref{sec:analysis}
describes data preparation and a simulation study in which the
methodologies were evaluated. Section \ref{sec:results} evaluates the
relative efficacy of DTWBMI as compared with Dynamic Time Warping Based
{[}Single{]} Imputation (DTWBI), interpolation, aggregation, Time Window
imputation and mean imputation. Finally in Section \ref{sec:conclusion},
we discuss relevant issues regarding the procedures presented in this
study.

\hypertarget{sec:background}{%
\section{Background}\label{sec:background}}

Dynamic Time Warping-Based Imputation (DTWBI) was developed for more
general use in the imputation of time series data. Because the method
makes use of patterns within the temporal characteristics of the data,
it is useful to evaluate its potential as a mechanism for correcting for
long gaps in trajectory data. In this section, we describe the building
blocks of this process.

\hypertarget{sec:metrics}{%
\subsection{Key metrics in travel data}\label{sec:metrics}}

Raw GPS data is comprised of location coordinates that can be used to
segment the day into stops and trips, and to calculate various metrics
of interest such as distance traveled. In comparison, travel diaries
lack this granularity of location information, usually containing only
start and end locations for trips. Travel diary data is therefore
limited to count, time, distance or travel mode statistics. Aggregation
often occurs at day- or week-level in these cases, and general
statistics estimated on the basis of the sample data. Location data
collected via a smartphone under ideal settings generate traces
representing the raw data of a person's location at a given point in
time. When the goal is to yield similar metrics, researchers must
process these raw data by calculating the distance, defining stops and
trips and establishing a mode of transportation.

Segmenting the trajectories into stops and trips without user input
requires selection of an algorithm. In this study, we use an algorithm
which defines stops on the basis of a radius and time parameter as
described in Montoliu et al. (2013). In this study, we apply Top-Down
Time Ratio segmentation on the raw data, summing the distances of the
segments (Meratnia and By, 2004). These two steps allow for the
calculation of count, time and distance statistics at the desired level
of aggregation.

\hypertarget{time-series}{%
\subsection{Time Series}\label{time-series}}

\begin{figure}
\includegraphics[scale=.8]{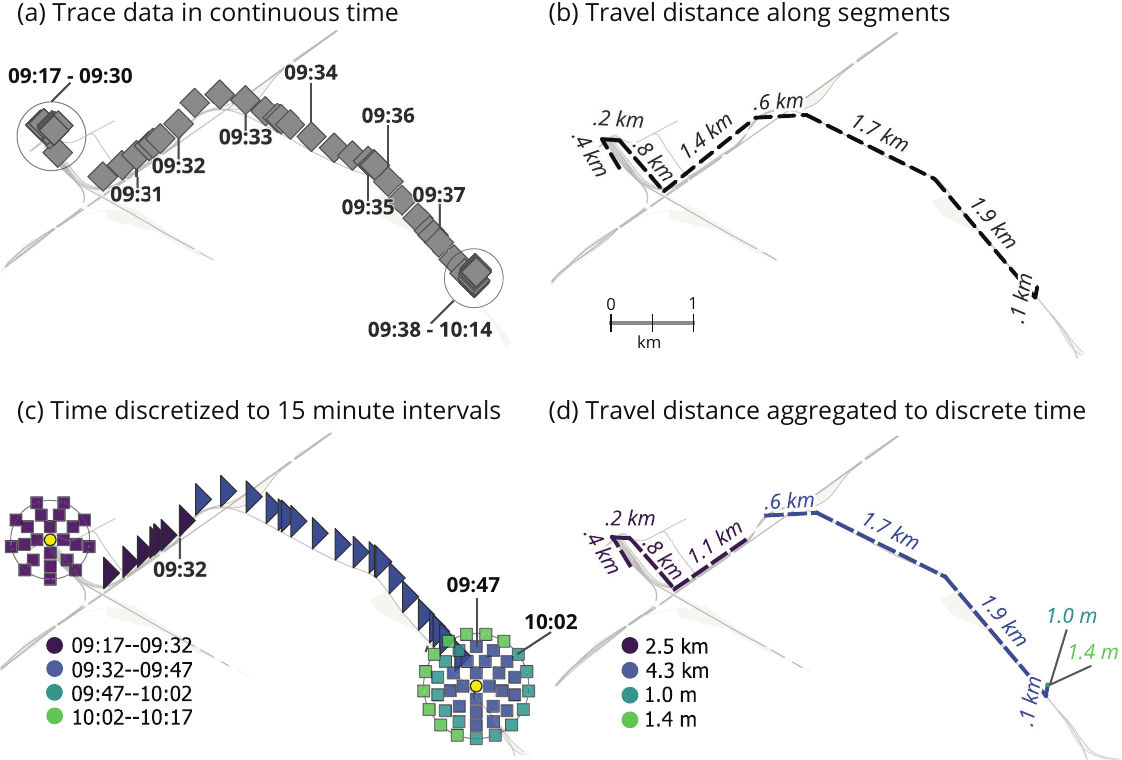}
\centering
\caption{Aggregation of travel distances into 15-minute intervals. (a) Timestampped geolocations with stay points (gray circles) and connecting track. (b) Segment-wise travel distance estimation. (c) Geolocations categorized into 15-minute segments, with stay points expanded for clarity; cutoffs are marked which occur during the track (09:32) and stay point (09:47, 10:02). (d) Segments recalculated within each time interval to sum travel distances, forming a discrete time series appropriate for our analyses.}
\label{fig:discretizingtime}
\end{figure}

It is possible to generalize individual trace data by considering it as
a time series of metrics of interest common between individuals. This
removes the specific geographic information associated with each
location. A person's travel behavior may be described as a time series
representing distances traveled over a discrete period or the increase
over time of the total covered area. Figure \ref{fig:discretizingtime}
shows an example in which trace data are discretized to 15 minute
intervals and distance is aggregated within these periods to produce a
four-element time series of distances. Such metrics can be calculated
across all users to describe aspects of travel behavior against which
different persons can be compared to find commonalities. This process
involves two important considerations, namely the selected metrics and
the way in which time is discretized.

Which metrics are selected should reflect the final analyses that the
researcher intends to employ. In this paper, we consider a selection of
key metrics typical to travel diaries, including travel distance (TD),
Radius of Gyration (RoG) and total stops (TS) as discussed in Section
\ref{sec:metrics}. This method is extensible to a wide variety of
metrics, including categorical measures such as transportation mode.

A secondary consideration involves the discretization of continuous
time. Metrics may be aggregated across intervals reflecting increments
of a given temporal resolution. As the temporal resolution changes, the
time series will lose detail which will decrease the potential
specificity in the matching process. The temporal resolution should be
of sufficient length to ensure that there are data within most
aggregated blocks containing complete data.

Using notation from Phan et al. (2020b), we describe a univariate time
series \(x\) as a series of \(N\) measurements or observations indexed
by (discrete) time \(t\), as shown in Equation \ref{eq:timeseries}

\begin{equation}
\label{eq:timeseries}
x = \{x_t|t = 1, 2, ..., N\}
\end{equation}

Using the example provided in Figure \ref{fig:discretizingtime}, we
would reduce this trace to a time series \(x\) comprised of \(N = 4\)
elements representing kilometers traveled during each time period \(t\):
\(\left\{2.5, 4.3, .001, .0014\right\}\).

In instances where \(x\) contains missing measurements, some time points
\(t\) will lack observations. We introduce a binary response variable
\(r_t\), indicating the presence (\(1\)) or absence (\(0\)) of data at
time \(t\). Missing data can manifest as isolated instances or span
multiple time points. For a given time index \(t\), if \(r_t = 0\) and
\(r_{t-1} = 1\), then \(t\) denotes the start of a gap. Let \(s\) be the
smallest index greater than \(t\) such that \(r_s = 1\), marking the
gap's end. The length of the gap \(T\) beginning at time \(t\) is then
given by Equation \ref{eq:gaplength}:

\begin{equation}
\label{eq:gaplength}
T^{t} = s - t
\end{equation}

\begin{figure}
\includegraphics[scale=.8]{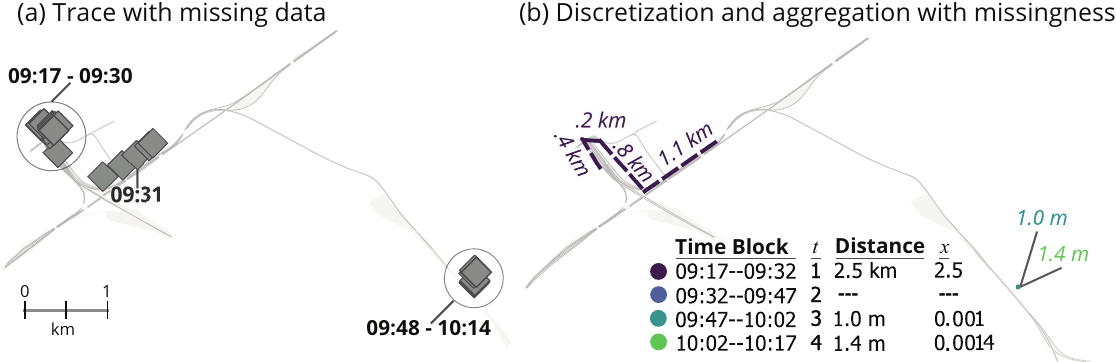}
\centering
\caption{Gaps between sucessive geolocations are considered to represent missing data if they exceed the length of the chosen discretization interval. (a) No locations are recorded after 09:31 and before 09:48, representing a gap of 16 minutes. (b) Time $t = 2$ therefore covers an unknown distance, and $x_2$ is missing.}
\label{fig:examplemissing}
\end{figure}

Figure \ref{fig:examplemissing} demonstrates the conversion of the
original trace to a time series \(x\) with missing data occurring during
time period \(t = 2\). In this case, the time series \(x_t\) would be
represented by \(\left\{2.5, -, .001, .0014\right\}\) and the response
series \(r_t\) by \(\left\{1, 0, 1, 1\right\}\). The length of the gap
\(T\) is here \(3 - 2 = 1\).

We extend both the time series \(x\) and the response variable \(r\) to
multiple persons indexed by \(k\), resulting in \(x^k\) and \(r^k\)
respectively. We can further generalize to a matrix \(X^k\) representing
multiple variables extracted as time series from an individual's data,
reflecting various aggregated metrics of interest, analogous to a design
matrix. As each series within a person retains the same missingness
pattern, the response vector \(r^k\) and associated gap lengths \(T\)
maintain the dimensionality of any contained \(x\).

\hypertarget{sec:dtw}{%
\subsection{Dynamic Time Warping}\label{sec:dtw}}

\begin{figure}
\includegraphics[scale=.7]{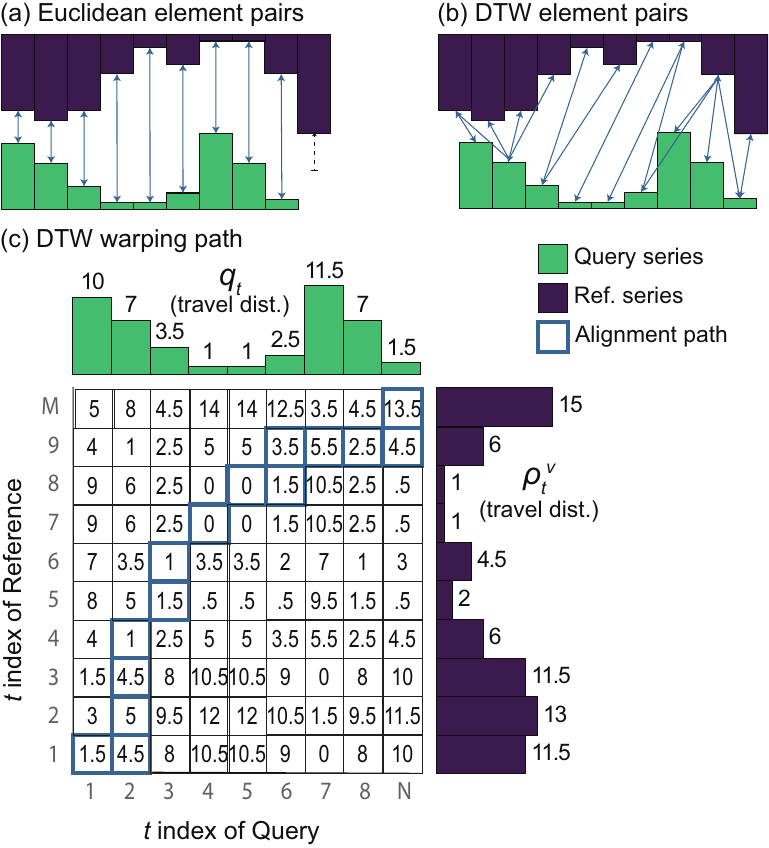}
\centering
\caption{Alignment of two time series representing kilometers traveled over a discretized interval of time. (a) Euclidean alignment, where each element in the query time series matches to the corresponding ordered element in the reference time series $\mathcal{R}$. (b) DTW alignment where the path of best fit is found where elements may match multiple times. (c) Cost matrix used to determine the past of best fit. The difference between each pair of elements is calculated and a path is drawn that results in the smallest total sum of differences, according to the standard DTW implementation in Equation 4.}
\label{fig:dtwwarp}
\end{figure}

In Dynamic Time Warping (DTW), we calculate a measure of similarity
between two time series as described in Equation \ref{eq:timeseries} by
finding a path of alignment between them that minimizes the sum of the
Euclidean distance (\(d(x_t,y_j) = x_t-y_j\)) between aligned element
pairs (Sakoe and Chiba, 1978). We distinguish between these series by
denoting the query series\footnote{Rabiner and Juang (1993) call this
  the test pattern \(\mathcal{T}\)} as \(q_t\) of length \(N\) and the
reference series\footnote{Rabiner and Juang (1993) use \(r\) to denote
  the singular instance, but this conflicts with our notation for the
  binary response variable.} as \(\rho_j\) of length \(M\). We call the
ordered selected pairs of points the warping path. Figure
\ref{fig:dtwwarp} compares a matching schema utilizing direct Euclidean
distance between ordered elements to a matching schema under DTW. If
\(M = N\), calculating the Euclidean distance as is shown in Figure
\ref{fig:dtwwarp}(a) can be done in the usual way, as shown in Equation
\ref{eq:euclideandist}

\begin{equation}
\label{eq:euclideandist}
dist(q,\rho) = ||q - \rho|| = \sqrt{(q_1 - \rho_1)^2 + (q_2-\rho_2)^2 + \dots + (q_N-\rho_N)^2}
\end{equation}

This requires both that \(q\) and \(\rho\) are the same length, and also
restricts the alignment to a one-to-one match with each \(q_t\) matching
to a \(\rho_j\) where \(t = j\). With DTW, we calculate a path of best
fit that minimizes the distance between the series by allowing a
disjoint matching of elements. A pairwise difference is calculated
between all indices \(t\) from \(1\) to \(N\) and all \(j\) from \(1\)
to \(M\), resulting in a matrix \(D\) where each element \(D_{tj}\) is
the difference between \(q_t\) and \(\rho_j\), as shown in Table
\ref{tab:dmat}.

\begin{longtable}[]{@{}
  >{\raggedleft\arraybackslash}p{(\columnwidth - 8\tabcolsep) * \real{0.1341}}
  >{\raggedleft\arraybackslash}p{(\columnwidth - 8\tabcolsep) * \real{0.2561}}
  >{\raggedleft\arraybackslash}p{(\columnwidth - 8\tabcolsep) * \real{0.2439}}
  >{\raggedleft\arraybackslash}p{(\columnwidth - 8\tabcolsep) * \real{0.1220}}
  >{\raggedleft\arraybackslash}p{(\columnwidth - 8\tabcolsep) * \real{0.2439}}@{}}
\caption{Distance matrix \(D\) representing the cost of all possible
routes \label{tab:dmat}}\tabularnewline
\toprule()
\begin{minipage}[b]{\linewidth}\raggedleft
\end{minipage} & \begin{minipage}[b]{\linewidth}\raggedleft
t = 1
\end{minipage} & \begin{minipage}[b]{\linewidth}\raggedleft
t = 2
\end{minipage} & \begin{minipage}[b]{\linewidth}\raggedleft
\(\cdots\)
\end{minipage} & \begin{minipage}[b]{\linewidth}\raggedleft
t = N
\end{minipage} \\
\midrule()
\endfirsthead
\toprule()
\begin{minipage}[b]{\linewidth}\raggedleft
\end{minipage} & \begin{minipage}[b]{\linewidth}\raggedleft
t = 1
\end{minipage} & \begin{minipage}[b]{\linewidth}\raggedleft
t = 2
\end{minipage} & \begin{minipage}[b]{\linewidth}\raggedleft
\(\cdots\)
\end{minipage} & \begin{minipage}[b]{\linewidth}\raggedleft
t = N
\end{minipage} \\
\midrule()
\endhead
j = 1 & \(dist(q_1, \rho_1)\) & \(dist(q_2,\rho_1)\) & \(\cdots\) &
\(dist(q_N,\rho_1)\) \\
j = 2 & \(dist(q_1,\rho_2)\) & \(dist(q_2,\rho_2)\) & \(\cdots\) &
\(dist(q_N,\rho_2)\) \\
\(\vdots\) & \(\vdots\) & \(\vdots\) & \(\ddots\) & \(\vdots\) \\
j = M & \(dist(q_1,\rho_M)\) & \(dist(q_2,\rho_M)\) & \(\cdots\) &
\(dist(q_N,\rho_M)\) \\
\bottomrule()
\end{longtable}

A path is selected through this matrix that provides the minimized cost
through all traversable paths as in Equation \ref{eq:mappath}, where
\(D(t,j)\) gives the minimum distance cost until point \(t\) and \(j\)
in the query and reference set respectively. A path may consist of
diagonal, vertical, or horizontal movements, reflected by
\(D(t-1,j-1)\), \(D(t-1,j)\), and \(D(t,j-1)\) respectively. Figure
\ref{fig:dtwwarp}(c) demonstrates an example calculation of the path of
best fit, given the cost matrix \(D\).

\begin{equation}
\label{eq:mappath}
D(t,j) = dist(t, j) + min{D(t-1,j-1),D(t-1,j),D(t,j-1)}
\end{equation}

In standard DTW, the first and last elements of query and reference
series must align to each other, meaning that the path must begin with
\(D(1,1)\) and end with \(D(M,N)\), regardless of cost. Additionally, a
path may allow no backtracking {[}\(D(t+1, ...)\) or \(D(..., j+1)\) are
not allowed{]}, and each element must be matched at least once {[}for
example, \(D(t-2, ...)\) or \(D(..., j-3)\) to skip over one element in
\(q\) or one element in \(\rho\) respectively{]}. In addition to these,
it may also be desirable to restrict the path in other ways, such as
requiring matched elements to be within a certain ordered distance, or
window, of each other.

\begin{figure}
\includegraphics[scale=.75]{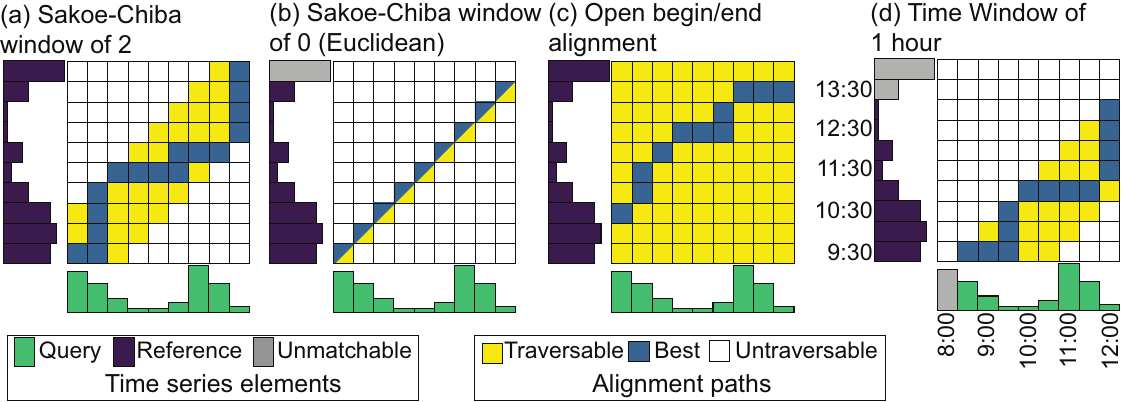}
\centering
\caption{Different alignment restrictions within Dynamic Time Warping. (a) Sakoe-Chiba window of 2, which allows alignment of elements at most two distant in order, (b) Sakoe-Chiba window of 0, where only one path exists in a one-to-one correspondence between query and reference, (c) easing of the beginning and ending restriction where first and last elements of the query need not align to the first and last elements of the reference vector, (d) is an example of the use of time windows reflecting true time, where the possible path is restricted to a one-hour window at each element of the query series.}
\label{fig:dtwrestrictions}
\end{figure}

Figure \ref{fig:dtwrestrictions} illustrates warping paths under
different constraints. Figure \ref{fig:dtwrestrictions}(a) adds the
constraint of the Sakoe-Chiba window of size 2. Differences are only
calculated between those elements falling within 2 elements of the query
series. In comparison, a Sakoe-Chiba window of size 0, shown in Figure
\ref{fig:dtwrestrictions}(b), allows only one path, matching each
element in the query to its ordered pair in the reference series.
Imposing this restriction requires that the query and reference series
are the same length, due to the restriction that the first and last
elements of the query and reference series must be aligned. Easing this
restriction is possible as well, as demonstrated in Figure
\ref{fig:dtwrestrictions}(c), with Open Begin/End DTW or subsequence
alignment (Tormene et al., 2009). All elements of the query vector must
align to at least one element of the reference vector, but the reverse
need not be true. When combined with a Sakoe-Chiba window of zero, this
is known as the best fitting subsequence search (Sakoe and Chiba, 1978).
Figure \ref{fig:dtwrestrictions}(d) illustrates our new method, which
restricts the available warping paths on the basis of a continuous-time
windowing parameter. Only elements which are within the given time
window are traversable by the algorithm.

Our objective is to determine the similarity between \(q\) and \(\rho\)
to evaluate the fitness of \(\rho\) to serve as a donor during a gap
\(T^t\) in \(q^k\). To this end, we define a collection of \(V\)
reference time series \(\mathcal{R}\), as shown in Equation
\ref{eq:refser}

\begin{equation}
\label{eq:refser}
\mathcal{R}^V = \{\rho^1, \rho^2, ..., \rho^V\}
\end{equation}

where each \(\rho^v\) is defined as in Equation \ref{eq:refserrho}

\begin{equation}
\label{eq:refserrho}
\rho^v = \{\rho^v_j|j = 1, 2, ..., M_v\}
\end{equation}

Table \ref{tab:mathcalr} provides an example of a univariate
\(\mathcal{R}\) containing aggregated distances in kilometers for
\(V = 3\) reference sets of varying lengths. If the only constraint
applied is a Sakoe-Chiba window of 0, requiring a one-to-one match, both
\(\rho^1\) and \(\rho^2\) can serve as reference series to the example
query from Figure \ref{fig:examplemissing} because \(M_v \ge N\), but
\(\rho^3\) cannot because it contains only 3 elements.

\begin{longtable}[]{@{}rrrrrr@{}}
\caption{\label{tab:mathcalr} \(\mathcal{R}\) with \(V = 3\) representing 3 distinct sets for
aligning on aggregated distance measurement across discrete time.
\(\rho^1\) is length \(M = 5\), and \(\rho^3\) is of length
\(M = 3\).}\tabularnewline
\toprule()
v & j = 1 & j = 2 & j = 3 & j = 4 & j = 5 \\
\midrule()
\endhead
1 & 2 km & 5 km & 0 km & 2 km & 5 km \\
2 & 0 km & 0 km & 0 km & 1 km & \\
3 & 1 km & 0 km & 2 km & & \\
\bottomrule()
\end{longtable}

Given that DTW can only compute distances for complete observations, we
split \(q\) into two segments:
\(q_{pre} = \{q_i | i = 1, 2, ..., t - 1\}\) and
\(q_{post} = \{q_i | i = s, s + 1, ..., N\}\), where \(t\) and \(s\)
mark the start and end of the gap respectively. For any given \(q^k\),
the associated \(\mathcal{R}\) may contain \(\rho\) from different
persons (\(v \neq k\)), or from the same person (\(v = k\)) but at times
not in \(q\) (\(j \notin i\)).

For each \(\rho^v\) in \(\mathcal{R}\), we calculate the distance
between each \(q_i \in \{q_{pre}, q_{post}\}\) and \(\rho_j\), find the
subset of \(j \in M\) that minimizes the total sum of distances, and sum
these distances to establish the overall dissimilarity, as shown in
Equation \ref{eq:sumdtw}:

\begin{equation}
\label{eq:sumdtw}
d(q, \rho^v) = \sum_i^N min(d(q_i, \rho_j))
\end{equation}

From our earlier example query where
\(q = \left\{2.5, -, .001, .0014\right\}\), we may create
\(q_{pre} = \{2.5\}\) and \(q_{post} = \{.001, .0014\}\). Table
\ref{tab:mathcalr} provides the example reference series
\(\rho^1 = \left\{2, 5, 0, 2, 5\right\}\). We similarly split \(\rho^1\)
at each permissible spot that will allow for a comparison against both
\(q_{pre}\) and \(q_{post}\), separated by a gap of length \(T = 1\),
generating the following two comparison reference series with the gap
from \(q\) occurring either at timepoint \(j = 2\) or \(j = 3\),
producing respectively \(\left\{2, (5), 0, 2\right\}\) and
\(\left\{5, (0), 2, 5\right\}\), in order to fit the pattern of
missingness of \(q\) (later allowing the distances \(5\) and \(0\)
respectively to be used to fill the gap). Table \ref{tab:qrcomp} shows
the comparison of both possible alignment moments of \(\rho^1\) with
this \(q\).

\begin{table}[]
\centering
\caption{All possible fits of $\mathcal{R}$ to $q$ under a Sakoe-Chiba window of 0. Respecting the length of $q_{pre}$, $T$, and $q_{post}$, $\rho^1$ has two possible alignments with its five elements, and $rho^2$ has only one possible alignment as it contains the same number of elements as $q$.  Column $T$ contains future potential donor values from each viable position in the $\rho^v$.}
\label{tab:qrcomp}
\begin{tabular}{@{}lrrrr@{}}
\toprule
                           & $q_{pre}$ & $T$ & \multicolumn{2}{c}{$q_{post}$} \\ \midrule
$q$                        & 2.5       & -   & .001          & .0014          \\
$\rho^1_{j\in\{1,2,3,4\}}$ & 2.0       & 5.0 & 0.0           & 2.0            \\
$\rho^1_{j\in\{2,3,4,5\}}$ & 5.0       & 0.0 & 2.0           & 5.0            \\ 
$\rho^2_{j\in\{1,2,3,4\}}$ & 0.0       & 0.0 & 0.0           & 1.0            \\
\bottomrule
\end{tabular}
\end{table}

The example alignment from Table \ref{tab:qrcomp} of \(q\) to
\(\rho^1_{j\in\{1,2,3,4\}}\) would produce a total dissimilarity of
\(|2.5 - 2| + |0.001 - 0| + |0.0014 - 2| = 2.4996\). Meanwhile, the
alignment of \(q\) to \(\rho^1_{j\in\{2,3,4,5\}}\) would produce a total
dissimilarity of \(|2.5 - 5| + |.001 - 2| + |0.0014 - 5| = 9.4976\), and
to \(\rho^2_{j\in\{1,2,3,4\}}\), a dissimilarity of
\(|2.5 - 0| + |0.001 - 0| + |0.0014 - 1| = 3.4996\). For
\(\mathcal{R}\), \(V\) total sets of indices and minimum dissimilarities
are calculated, representing the best fit for each \(\rho^v\). Because
\(\rho^1_{j\in\{1,2,3,4\}}\) is a better fit than
\(\rho^1_{j\in\{2,3,4,5\}}\), the alignment position and total
dissimilarity of \(2.4996\) would be returned for \(\rho^1\). Because
only one possible alignment exists for \(\rho^2\), alignment position
\(\{1,2,3,4\}\) and dissimilarity \(3.4996\) will be returned for
\(\rho^2\).

We may constrain \(\mathcal{R}\) by applying a set of restrictions,
which we define as the function \(\phi\), as shown in Equation
\ref{eq:phi}. This may limit persons (\(v\)) or time (\(j\)), or
restrict the set of allowable minimization paths, as in Equation
\ref{eq:sumdtw}.

\begin{equation}
\label{eq:phi}
d(x,r^j) = min(d_\phi(x,r^j))
\end{equation}

One logical limitation is implementing a time difference threshold
between the query and reference sets. The maximum allowed difference in
time between points in the query and points in the reference set we
refer to as the window parameter. This limits the possible paths of
alignment to be temporally similar so that morning travel behavior is
imputed in the morning, midday travel behavior imputed for the midday,
etc. Under this limitation, \(\phi\) imposes the restriction
\(|t_i - t_j| \le window\). This type of temporal windowing can also be
applied to apply to days of the week or to allow for weekend/weekday
restrictions. Reducing the number of computed similarities also has the
side effect of decreasing the total computational time.

It is unclear which limitations perform best in this situation. In
practice, one needs to make a choice for the parameters in \(\phi\), and
it is not obvious how this should be done, nor have earlier studies
provided much guidance. For this reason, we implemented a simulation
study in the context of a week-long travel diary study in which GPS
locations were measured over a period of a week to illustrate how
different choices lead to different matching sets, and affect the
statistics of interest. \protect\hyperlink{sec:appendix}{Appendix A}
describes this simulation study.

\hypertarget{multiple-imputation}{%
\subsection{Multiple Imputation}\label{multiple-imputation}}

It is quite likely that the DTW procedure as explained in the previous
section yields several possible matches. This is desirable, as it allows
for the selection of a number of different matches instead of a single
``best-fitting'' example. We select \(m\) matches to create \(m\)
different data sets against which we will calculate final statistics in
a process called Multiple Imputation (Rubin, 2004). The basic premise
behind multiple imputation is that the creation of multiple complete
datasets from a single incomplete dataset allows us to obtain standard
errors that are appropriately large (Buuren, 2018). This is valuable
because it permits calculation of uncertainty.

Multiple imputation can be used to describe a range of methods that can
be used to fill in missingness. A subset of these methods, among which
are Hot Deck imputation and Predictive Mean Matching, describe
techniques that impute missing values with observed values (Little,
1988). This provides more plausible values as they are drawn from true
occurrences, with the added advantage of requiring little in the way of
model specification. Implementation of this method requires generation
of a particular number of candidates on some basis of similarity. This
similarity between the query and the reference sets that are possible
donors is used to generate a probability of selection for each reference
set (Siddique and Belin, 2008). This can be tuned by adjusting a
parameter in the calculation to increase or decrease the probability of
selection for similar donors. The multiple imputation procedure can be
chained such that data that are incomplete at the start of the
imputation procedure may themselves be used to impute subsequent data
(Van Buuren and Oudshoorn, 2000). This improves the quality of the
imputations and in the case of missing time series data, allows for all
available data to be used when determining the set of closest matches.

\hypertarget{dynamic-time-warping-based-imputation}{%
\subsection{Dynamic Time Warping-Based
Imputation}\label{dynamic-time-warping-based-imputation}}

\begin{figure}
\includegraphics[scale=1]{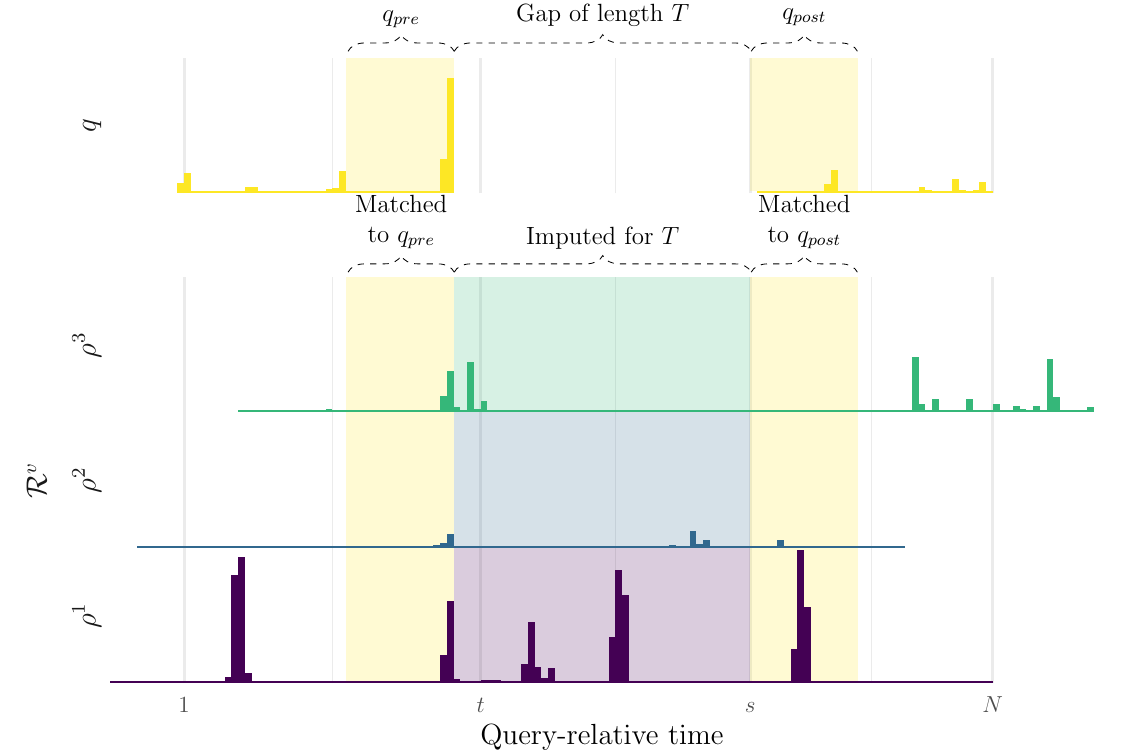}
\centering
\caption{An example of Dynamic Time Warping-Based Imputation. The match buffer for the query $q$ is selected surrounding the gap, creating $q_{pre}$ and $q_{post}$. Artificial gaps of length $T$ are created in each reference series $\rho$,  and buffers equal in length to $q_{pre}$ and $q_{post}$ (highlighted in yellow) are selected for matching against the query. Selected candidates donate the data between the buffer elements (highlighted in green, blue and purple respectively). The differing lengths of the series illustrate DTW in practice and its capacity to start at different points.}
\label{fig:dtwbi}
\end{figure}

The implementation of DTW as a selection mechanism for imputation has
been previously described (Phan et al., 2020a, 2018a, 2018b). DTW may be
used to find the best-fitting subsequence within candidate series based
on matches against the observed data immediately preceding and following
a gap in a target series in a method called Dynamic Time Warping-Based
Imputation (DTWBI). Variations on this concept have been implemented as
a selection mechanism for gap-filling in various disciplines (Hung et
al., 2022; Kostadinova et al., 2012; Zhang et al., 2019). This paper
differs somewhat to previous implementations of DTWBI and extends the
single imputation to multiple imputation.

First, the restrictions on the start and stop of query alignment are
relaxed as in Open Begin/End DTW. Second, a strict one-to-one
concordance is imposed between query and reference time series, which
enforces the identifiability of the matching sequence and makes more
intuitive sense with discretized time series. We also introduce Time
Window, which is similar to the Sakoe-Chiba window, but restricts the
query alignment to reference time points occurring within a specified
time difference. This is done in order to align travel behaviors only
with other travel behaviors occurring at the same time of day (although
the days may differ). Lastly, we replace the gap in the target time
series with a single time series element representing the interpolated
measure, then excise and interpolate along each potential candidate at
each potential gap moment, allowing us to make full use of all data.
Figure \ref{fig:dtwbi} provides an example this procedure.

DTWBI has many parameters that describe its implementation. The efficacy
of DTWBI is expected to depend on the selection of these parameters. In
these analyses, we vary four parameters: two impacting the mechanisms of
the dynamic time warping alignment, and two impacting imputation
procedure. The parameters match buffer and time window reflect integral
considerations for DTW alignment, while the parameters candidate
specificity and number of imputations impact the subsequent imputation
procedure.

The match buffer parameter describes the number of elements considered
before and after the gap. A buffer of zero elements is equivalent to
selecting candidates independent of DTW similarity. A larger buffer
prefers longer patterns such as a person's work commute while a shorter
buffer prefers shorter patterns, such as a person's moving preceding a
gap but not afterwards.

The time window parameter describes the maximum time difference
allowable between matchable elements in the target and query time
series. For example, a window of two hours would allow a query element
of 8AM to match against target elements occurring between 6AM and 10AM.
The tighter the time window, the more candidate matches are restricted
to conform to time-based routines. The wider the time window, the more
candidate matches are allowed to favor similar behavior without
requiring the time element. A time window of 12 hours would allow for
unrestricted pattern matching in the time series (12 hours before, and
12 hours afterward).

DTWBMI is the extension of DTWBI to the multiple imputation framework.
Whereas in DTWBI, only the best fitting candidate is selected for
imputation, with DTWBMI we select a number of different candidates. This
adds new parameters that can impact the imputation -- the number of
imputations and the candidate specificity. A higher number of
imputations means that we will average results out across a number of
worse- and better-fitting candidates from whose data the imputed data
will be drawn. This has the advantage of allowing us to quantify the
uncertainty of each estimate. Candidate specificity is a parameter that
describes the relationship between the selection probability of an
imputation candidate and how close a match the donor is. A higher
candidate specificity leads to a higher preference for similarity,
versus a low candidate specificity which gives us a high tolerance for
misspecification. If the candidate specificity is too high, we may
select the same donor across all imputations, reducing to the DTWBI case
of a single best donor. If the candidate specificity is too low, we may
select donors without regard for similarity at all, reducing to the Time
Window imputation case.

These four parameters produce a number of different combinations whose
effectiveness is expected to depend not only on the interactions with
the other parameters, but also on differences in the data. Of primary
interest is selecting parameters sets that are appropriate to extent of
the information available within the data. As the amount of data
increases, either through increasing the total number of persons or the
length of data collection, a longer match buffer will be beneficial in
increasing the chance that identical routes will be identified during
the alignment procedure. Similarly, a higher candidate specificity would
be beneficial in a high-information setting, increasing the frequency of
selection of very-similar imputation candidates. The expected impact of
the parameters time window and number of imputations is less clear. A
set of parameters more appropriate for a low-information data set in
which there is little overlapping data either within or between
individuals may reflect instantaneous travel patterns, such as travel
immediately preceding or following a gap, or may have a shorter time
window in order to benefit from the temporal aspects of travel behavior.
A low-information parameter set may thus prefer a shorter time window,
shorter match buffer, reduced candidate specificity, and an increased
number of imputation streams.

We expect improved performance of the high-information variant relative
to the low-information variant to be related to the type of travel
behavior generated by a person, but restricted by the amount of
available data for the person. Persons with very consistent travel
patterns across days would be expected to benefit from a
high-information variant as they will be more likely to match against
their own near-identical travel behavior than against that of others.
Persons with a varying travel pattern across days, but whose travel
behavior is unlikely to deviate from typical travel patterns within the
sample, are unlikely to benefit from the high-information variant until
the study periods are long enough to encapsulate most of the deviation.
A third type of person, for whom travel patterns vary over the days, but
whose activities deviate from the travel patterns of others within the
sample (e.g.~a large number of stops/tracks, long travels), may find
that the low-information variant performs acceptably in shorter gap, but
poorly in longer gaps, and that a high-information variant would have
poorer performance until a sufficient number of similar travel
activities were observed for the same or other individuals.

We performed a simulation study to establish these two sets of
parameters, leading to two methods called DTWBMI-HI and DTWBMI-LO,
representing high- and low-information availability respectively. Full
results are available in \protect\hyperlink{sec:appendix}{Appendix A}.

\hypertarget{sec:analysis}{%
\section{Analysis strategy}\label{sec:analysis}}

This section describes the selection of a suitable data set for the
simulation study, criteria for comparison, and the other methods against
which this new method is to be compared.

\hypertarget{sec:exdata}{%
\subsection{Example data set}\label{sec:exdata}}

As a motivating example, we consider the data collected from a 2018
field test of the Statistics Netherlands travel app. This field test
concerned 1902 sample persons aged 16 and older. ODiN is an online-only
study of individual mobility in the Dutch population (Centraal Bureau
voor de Statistiek, 2022). Both groups of respondents were contacted via
post with a request to download the application onto their personal
mobile devices, register using the enclosed personal username and
password, and record seven days of movement behavior. Full details on
app methodology and data structure have been outlined in McCool et al.
(2021). The app captured a participant's location once per second while
the person was determined to be in motion, and once per minute while the
person was determined to be stationary. This determination was based
upon an algorithm that assessed whether or not the displacement between
recorded intervals exceeded thresholds indicating movement behavior.
Collectively, a total of 2087 person days were recorded amongst 576
participants.

We describe the missing data in terms of covered time and with respect
to the number and length of gaps. A natural gap occurs between each two
successive recorded locations within a single person's trajectory. If
this gap is very small, on the order of seconds, very little information
on a person's continuous trajectory is lost. As the time elapsed between
locations increases, the amount of potential information lost increases.
Deciding on a maximum allowable gap length is somewhat arbitrary, but
should depend on the smallest movement behavior of interest to the
researcher. The missing data patterns in this data set were determined
on the basis of a maximum gap time of six minutes. The contiguous length
of time elapsing for a person without exceeding the maximum gap time
between any two successive locations we refer to as the covered time.
Covered time can be expressed as a percentage of a discrete length of
time, such as an hour or a day, as a measure of data completeness. We
can then speak of the hourly, daily coverage. In this data set, the
average length of covered time in a day was 8.2 hours, corresponding to
a mean coverage of 34.2\%. The mean number of gaps per calendar day
across all participants was 4.0, and the mean gap length was 2.23 hours
in length.

\begin{figure}
\centering
\includegraphics{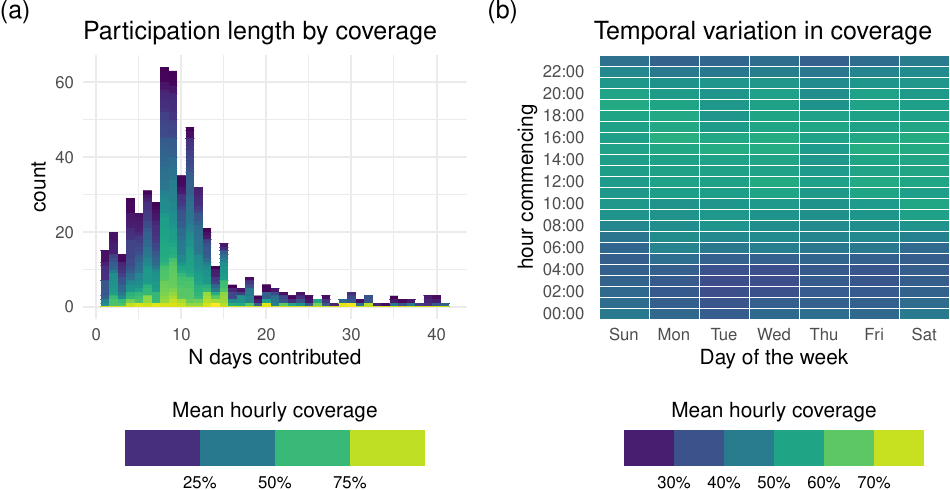}
\caption{Increased length of participation may increase the value of
high-information gap-filling mechanisms by providing more reference sets
of a person's own behavior. Missingness that occurrs in the evening has
a low likelihood of containing travel behavior, which may reduce the
impact of high-information gap filling mechanisms.
\label{fig:missingnessfig}}
\end{figure}

Following from the initial missing data analyses, we identified two
specific aspects of interest across which we expect the effectiveness of
the gap-filling methods to differ: participation length, and temporal
variation. First, participants recorded data on a variable number of
days, ranging from one to 43, creating a natural variation in the total
information available per individual. Figure \ref{fig:missingnessfig}
(a) shows the distribution of persons in the data set by the number of
days on which they submitted at least some data, grouped by their
average hourly coverage. Some respondents maintain high levels of daily
coverage, represented by the brighter yellow color, over a long period
of time. As the amount of data from a person's own travel behavior
increases, the additional days become available to form reference sets
for imputing potentially missing data. Secondly, McCool et al. (2021)
noted previously that nighttime hours have an increased incidence of
missingness, due in part to operating system behaviors that restrict
processing when the device is not in use. Figure
\ref{fig:missingnessfig} (b) illustrates the average coverage across
time and day of the week. Coverage is lowest during night time hours and
highest during commuting times through the week. Patterns of coverage
are more dispersed within the weekends. Because most travel behavior
occurs during the daytime hours, and much of the missing data occurs at
night, imputation of missing data occurring during the evening is
unlikely to outperform simpler methods.

\hypertarget{comparison}{%
\subsection{Comparison}\label{comparison}}

\begin{figure}
\includegraphics[scale=.8]{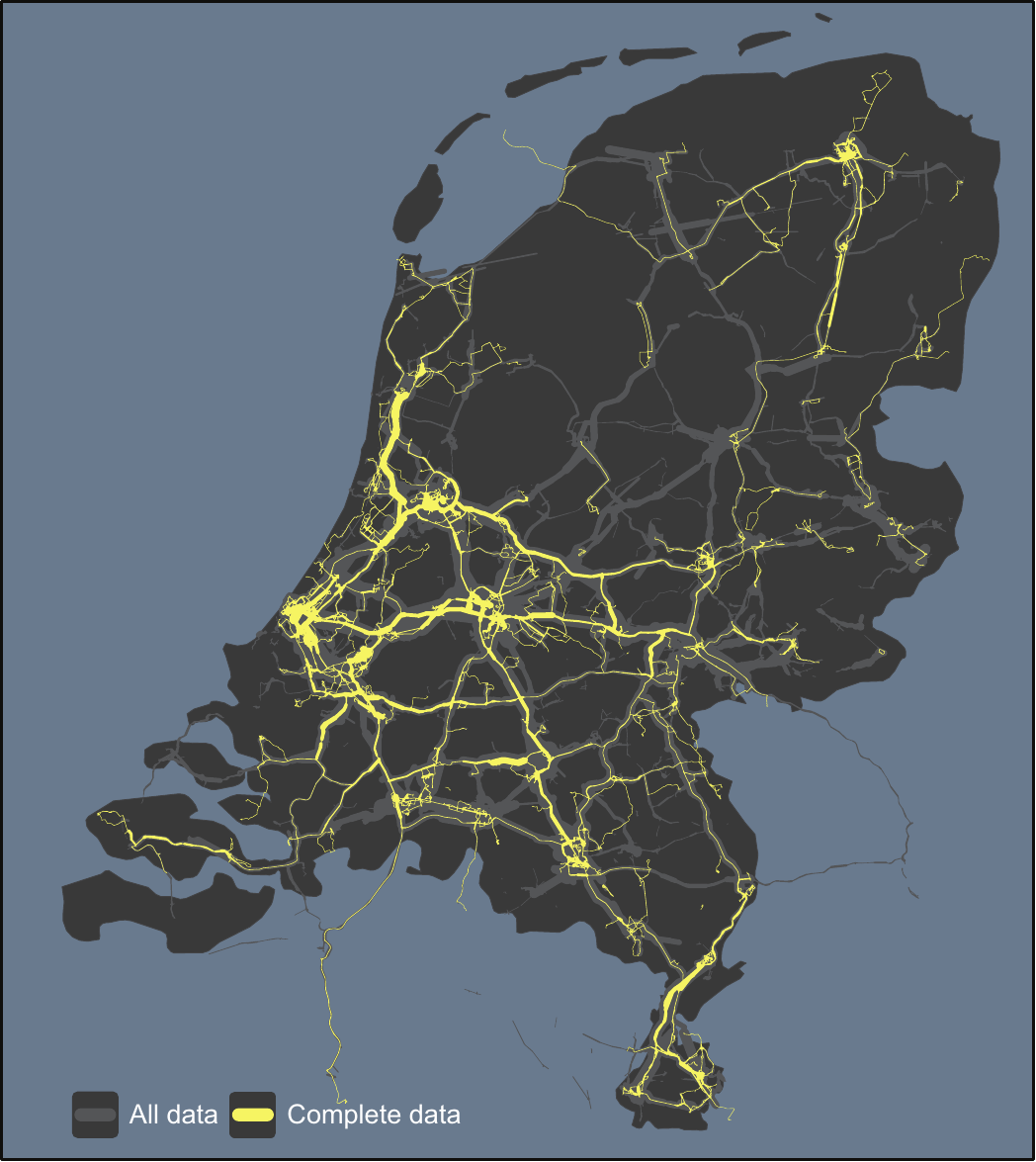}
\centering
\caption{Spatial density of trajectories recorded in the app by participants. Light grey areas show segments represented in the full set of data, which includes both data sets with gaps as well as data sets covering less than 24 hours in total. Yellow areas show the set of data used in the simulation study, which required at least 24 hours of contiguous data with no gaps exceeding 6 minutes.}
\caption*{Note: Densities are calculated on the basis of the number of intersections with segments from other users, Non-intersecting segments were removed to protect privacy, but were used in the simulation study.}
\label{fig:trajmap}
\end{figure}

From the data set described in Section \ref{sec:exdata}, we selected all
contiguous periods of at least 24 hours and with no gaps exceeding 6
minutes in duration. Individual trajectories with implausible speeds
between successive locations were manually inspected to identify which
locations were more likely to be incorrect. These were then flagged for
removal, and the data were again filtered to remove any contiguous sets
with gaps exceeding 6 minutes. Following this, 274 contiguous sets
across 143 respondents remained. Figure \ref{fig:trajmap} shows the
spatial density of all recorded trajectories relative to the density of
the set of trajectories covering at least 24 hours.

Because the length of the gap plays an integral role in the accuracy
with which the data can be imputed, we created five different
missingness scenarios, representing gap lengths of one hour, three
hours, six hours, 10 hours and 12 hours. Within each scenario, 100 of
the 274 sets were selected at random for introduction of missingness.
The missingness was introduced into each of the selected sets as a
single contiguous period. Each scenario randomized which sets contained
missing data, as well as the start time of the missing data, but all
methods compared within each scenario were applied to the same data.

In total, we compare six different methods on their capacity to address
missing data induced at various levels. These methods are: 1) linear
interpolation (LI), 2) mean imputation (MI), 3) Time Window Imputation
(TWI), 4) Dynamic Time Warping-Based {[}Single{]} Imputation (DTWBI), 5)
Dynamic Time Warping-Based Multiple Imputation under High Information
(DTWBMI-HI), and 6) Dynamic Time Warping-Based Multiple Imputation under
Low Information (DTWBMI-LO). All methods were implemented in the R
language and are available at
\url{https://anonymous.4open.science/r/DTWBI_Analyses-0178}.

\textbf{LI} calculates the distance between the last point before the
gap and the first point following the gap using the Haversine formula.
This distance is divided equally across the N discrete missingness
periods. The gap is then filled with a time series of identical travel
behavior. This is done in order to preserve the time series format.

\textbf{MI} filled each gap with the mean value per hour of the
applicable statistic, calculated on the basis of the non-missing
portion. This personal mean is used to fill the gap, creating a time
series of the travel behavior equal to the length of the gap.

\textbf{TWI} imputes missing data comparable time window selected from
candidates with complete data. Candidates' travel behavior similarity is
not otherwise assessed. Both the one-hour and three-hour values were
considered for the time window parameter with similar performance. We
compare one-hour TWI because of its slightly better aggregate
performance. Ten imputations were performed because of the theoretical
value of increasing sufficient variability.

\textbf{DTWBI} selects the best-fitting candidate on the basis of DTW
with a match buffer of eight hours and a time window of one hour. This
parameter set was selected on a theoretical basis for a desirable
performance assuming one near-identical travel pattern was available as
an imputation candidate.

\textbf{DTWBMI-HI} selects the three best-fitting candidates based on
DTW alignment, utilizing an extended match buffer of eight hours, and a
high candidate specificity to target finding a few very close matches --
such as a person's own activity trajectory when available. The
parameters for match buffer and candidate specificity were theoretically
posited to enhance matching precision, selecting for trajectories that
had close alignment over a longer interval. The simulation study
described in Appendix \ref{sec:appendix} demonstrated the viability of
the theoretical combination, and suggested a 12-hour time window and
three imputation candidates as complimentary on the basis of providing
the lowest absolute bias as well as an acceptable profile across other
performance measures. Both of these align with theory, as we are likely
to have fewer very close matches and will thus prefer fewer candidates,
and the lack of time restriction may allow for matching morning commutes
against evening commutes, or for otherwise identical trajectories that
are made at different points in time.

\textbf{DTWBMI-LO} selects the 10 best fitting candidates on the basis
of DTW alignment, using a match buffer of one hour, a medium candidate
specificity, and a time window of three hours. In the case of DTWBMI-LO,
the simulation study detailed in Appendix A was used as the primary
basis for parameter selection. This combination provided the overall
best profile across the selected measures of performance. Subsequent
investigation of the matches produced by this method suggested a
theoretical mechanism of action relying on a general relationship
between travel behavior immediately preceding/following a gap, and the
travel behavior contained within the gap.

Results are split according to gap length, and also with respect to the
two characteristics identified in \ref{sec:exdata}: number of available
sets and night-only missingness. This may allow for fine-tuning method
selection.

\hypertarget{sec:perfcrit}{%
\subsection{Performance criteria}\label{sec:perfcrit}}

We selected the following two key travel metrics for evaluation of
performance: total distance and number of stops. We compare the
parameters on the basis of Root Mean Square Error (RMSE) and mean
absolute bias (Bias), as well as on a set of metrics developed to assess
the accuracy and directional bias of the imputed travel distance and
number of periods spent moving. RMSE and Bias both assess the accuracy
of the underlying imputed metric in absolute terms. Because different
parameter sets generated either a significant upward bias or downward
bias on total distance, we compared under- and overestimation
separately.

Travel Periods Overestimated (TP over) reflects the percentage of
15-minute travel periods imputed that did not exist in the true data
set. Conversely, Travel Periods Underestimated (TP Under) reflects the
percentage of the true number of periods that were spent in movement
that were not reflected in the imputation. Travel Period Accuracy (TP
Acc.) reflects the percentage agreement with the total count of
moving/stationary periods between the true data and the imputed data.

Distance overestimated (Dist Over) reflects only the upward bias of the
imputed values relative to the true distance. Distance Underestimated
(Dist Under) similarly reflects only the downward bias of the imputed
values relative to the true distance. Both are expressed in kilometers.

\hypertarget{sec:results}{%
\section{Results}\label{sec:results}}

\hypertarget{travel-distance}{%
\subsection{Travel distance}\label{travel-distance}}

Table 1 shows results for travel distance imputation following
imputation with the six tested methods: linear interpolation, mean
imputation, TWI, DTWBI, DTWBMI-HI, and DTWBMI-LO, averaged across all
five missingness scenarios. The best average performance in terms of
absolute bias (Abs Bias) is DTWBMI-LO, with a mean bias of 0.6 Km. The
median bias (Med Bias) for all DTW methods is less than 100 meters, and
300 meters for the linear interpolation method. We can break out
absolute bias into distance overestimated (Dist Over) or underestimated
(Dist Under) which allows investigation of systematic directional biases
in the gap-filling procedure. Linear interpolation is incapable of
overestimating the travel distance -- instead all bias reflects an
underestimation of travel distance of 5.8Km on average. All DTW-based
methods systematically underestimate travel distance. DTWBMI-LO
outperforms the other methods; despite a slight underestimation of
travel distance, both overestimation and underestimation are small
relative to other methods, and the difference between the average
overestimation and underestimation is small.

The travel period metrics offer insight into the shape of the
gap-filling method. DTWBI provides the highest accuracy with respect to
reproducing the correct number of 15-minute periods in which a person
was traveling. DTWBMI-HI and DTWBMI-LO also offer approximately 95\%
accuracy. TWI and linear interpolation are less accurate than the other
methods. Travel period overestimation (TP Over) and underestimation (TP
Under) allow for investigation of systematic biases in travel period
estimation. DTWBMI-LO is the only method for which underestimation and
overestimation percentages are similar and also low. Mean imputation has
an overall TP Over of zero because the mean value imputed for the
missing values does not exceed the movement threshold. Linear
interpolation, although it is not capable of exceeding the total
traveled distance, can assign a distance exceeding the movement
threshold to a period which originally contained no movement. Here,
DTWBI demonstrates a problem with systematic underestimation of travel
periods.

\begin{longtable}{l|rrrrrrrr}
\caption*{
{\large Table 1: Method comparison across all cases} \\ 
{\small Imputing travel distance}
} \\ 
\toprule
\multicolumn{1}{l}{} & Abs Bias & Med Bias & Dist Over & Dist Under & RMSE & TP Acc. & TP Over & TP Under \\ 
\midrule\addlinespace[2.5pt]
LI & $5.9$ Km & $-0.3$ Km & $0.0$ Km & $5.9$ Km & $1.05$ & $91.9\%$ & $3.7\%$ & $36.8\%$ \\ 
MI & $1.9$ Km & $5.7$ Km & $7.8$ Km & $5.8$ Km & $1.38$ & $93.7\%$ & $0.0\%$ & $42.4\%$ \\ 
TWI & $1.1$ Km & $2.0$ Km & $8.4$ Km & $7.3$ Km & $1.97$ & $91.8\%$ & $23.5\%$ & $23.3\%$ \\ 
DTWBI & $1.8$ Km & $0.0$ Km & $2.7$ Km & $4.5$ Km & $1.40$ & $95.5\%$ & $7.4\%$ & $20.8\%$ \\ 
DTWBMI-HI & $0.7$ Km & $0.0$ Km & $4.8$ Km & $5.5$ Km & $1.49$ & $94.2\%$ & $13.4\%$ & $21.3\%$ \\ 
DTWBMI-LO & $0.6$ Km & $0.0$ Km & $3.2$ Km & $3.8$ Km & $1.44$ & $95.3\%$ & $12.2\%$ & $16.1\%$ \\ 
\bottomrule
\end{longtable}

Previous studies have demonstrated that the length of the gap is an
important consideration in deciding on a method. Table 2 shows method
performance metrics across the five simulations with differing gap
lengths. With a gap one hour in length, DTWBI, DTWBMI-LO and mean
imputation provided the least bias on average. Mean imputation offers
quite a high median bias at 1.9 Km on average, with other methods all
less than 100 meters. The over- and underestimation was relatively low
across all methods due to the small likelihood of one hour intervals
containing travel over long distances. The performance of DTWBI and
DTWBMI-LO are very similar, performing best on these metrics, containing
over- and underestimation in a similar ratio and to a small degree. All
travel period metrics also favor DTWBMI-LO, although linear
interpolation also offers acceptable performance.

As the gap length increases to 3 hours, the performance of DTWBI and
DTWBMI-LO remain sufficient. Although both offer favorable profiles with
low bias, little evidence for large systematic bias in distance
estimation. LI and DTWBMI-HI offer acceptable performance with regard to
overall bias metrics, but performs less well in the assessment of the
imputed shape and accuracy of the travel periods. Methods TWI and mean
imputation do not offer favorable profiles in the 3-hour gap length
condition. In the 6-hour gap condition, DTWBMI-LO demonstrates the best
performance. While TWI has the lowest absolute bias, it offers worse
performance across the other metrics. DTWBMI-LO has a more favorable
shape profile, with both 14\% over and underestimation of the travel
periods. The other methods offer relatively worse performance on
average. Within the 10-hour gap simulation, DTWBMI-LO offers the best
performance across all methods. The absolute bias is 300 meters on
average, although the bias split by over- and underestimation is 6.8 Km
and 6.5 Km respectively. Here, DTWBMI-HI also offers an acceptable
performance with a small bias of 1.1 Km, and a favorable shape profile.
In the 12-hour gap simulation, single imputation by DTWBI offers the
best profile on average, with a bias of only 100 meters, although there
is a meaningful bias towards travel period underestimation. DTWBMI-LO
performs well here as well, with an absolute bias of only 200 meters,
roughly equitable travel period over- and underestimation, but a median
bias of 1.7 Km. Overall, increased gap length tends towards increased
absolute bias across most methods, and an increase in both
overestimation and underestimation of the total distance and the number
of travel periods. The median bias remains low for all DTW-based
methods, even as gap length increases to 12 hours. Both DTWBMI-LO and
DTWBI are able to provide much better approximations of travel behavior
than linear interpolation at gap lengths of 10 and 12 hours.

\begin{longtable}{l|rrrrrrrr}
\caption*{
{\large Table 2: Method comparison across gap length}
} \\ 
\toprule
\multicolumn{1}{l}{} & Abs Bias & Med Bias & Dist Over & Dist Under & RMSE & TP Acc. & TP Over & TP Under \\ 
\midrule\addlinespace[2.5pt]
\multicolumn{9}{l}{1 hr} \\ 
\midrule\addlinespace[2.5pt]
LI & $0.8$ Km & $0.0$ Km & $0.0$ Km & $0.8$ Km & $0.79$ & $93.0\%$ & $5.5\%$ & $5.0\%$ \\ 
MI & $0.9$ Km & $1.9$ Km & $1.5$ Km & $2.5$ Km & $1.37$ & $93.0\%$ & $0.0\%$ & $16.0\%$ \\ 
TWI & $1.4$ Km & $0.2$ Km & $1.2$ Km & $2.6$ Km & $1.47$ & $89.3\%$ & $10.7\%$ & $13.4\%$ \\ 
DTWBI & $0.5$ Km & $0.0$ Km & $0.3$ Km & $0.9$ Km & $0.96$ & $95.0\%$ & $3.2\%$ & $8.8\%$ \\ 
DTWBMI-HI & $1.4$ Km & $0.0$ Km & $0.1$ Km & $1.6$ Km & $0.88$ & $94.1\%$ & $3.1\%$ & $10.3\%$ \\ 
DTWBMI-LO & $0.7$ Km & $0.0$ Km & $0.2$ Km & $0.9$ Km & $0.75$ & $95.7\%$ & $3.2\%$ & $5.5\%$ \\ 
\midrule\addlinespace[2.5pt]
\multicolumn{9}{l}{3 hrs} \\ 
\midrule\addlinespace[2.5pt]
LI & $2.4$ Km & $-0.1$ Km & $0.0$ Km & $2.4$ Km & $0.76$ & $92.0\%$ & $2.5\%$ & $28.0\%$ \\ 
MI & $1.2$ Km & $5.7$ Km & $4.2$ Km & $3.0$ Km & $1.08$ & $93.0\%$ & $0.0\%$ & $32.0\%$ \\ 
TWI & $1.9$ Km & $1.6$ Km & $5.4$ Km & $3.5$ Km & $1.69$ & $90.4\%$ & $18.7\%$ & $20.8\%$ \\ 
DTWBI & $1.0$ Km & $0.0$ Km & $1.0$ Km & $2.0$ Km & $1.06$ & $94.5\%$ & $6.2\%$ & $16.7\%$ \\ 
DTWBMI-HI & $1.9$ Km & $0.0$ Km & $1.1$ Km & $3.0$ Km & $1.05$ & $93.4\%$ & $8.2\%$ & $19.7\%$ \\ 
DTWBMI-LO & $0.9$ Km & $0.0$ Km & $0.8$ Km & $1.7$ Km & $1.04$ & $94.5\%$ & $7.7\%$ & $13.9\%$ \\ 
\midrule\addlinespace[2.5pt]
\multicolumn{9}{l}{6 hrs} \\ 
\midrule\addlinespace[2.5pt]
LI & $5.4$ Km & $-0.2$ Km & $0.0$ Km & $5.4$ Km & $0.99$ & $92.9\%$ & $3.3\%$ & $37.0\%$ \\ 
MI & $1.4$ Km & $11.5$ Km & $7.9$ Km & $6.4$ Km & $1.31$ & $94.5\%$ & $0.0\%$ & $42.0\%$ \\ 
TWI & $0.2$ Km & $3.3$ Km & $7.6$ Km & $7.8$ Km & $1.89$ & $93.0\%$ & $24.6\%$ & $22.7\%$ \\ 
DTWBI & $3.4$ Km & $0.0$ Km & $0.9$ Km & $4.3$ Km & $1.37$ & $96.5\%$ & $7.4\%$ & $17.6\%$ \\ 
DTWBMI-HI & $3.4$ Km & $0.1$ Km & $2.9$ Km & $6.3$ Km & $1.44$ & $94.8\%$ & $11.8\%$ & $20.1\%$ \\ 
DTWBMI-LO & $1.9$ Km & $0.1$ Km & $2.3$ Km & $4.1$ Km & $1.42$ & $95.6\%$ & $13.1\%$ & $14.8\%$ \\ 
\midrule\addlinespace[2.5pt]
\multicolumn{9}{l}{10 hrs} \\ 
\midrule\addlinespace[2.5pt]
LI & $11.7$ Km & $-5.1$ Km & $0.0$ Km & $11.7$ Km & $1.76$ & $87.5\%$ & $6.2\%$ & $58.0\%$ \\ 
MI & $2.9$ Km & $8.9$ Km & $9.0$ Km & $11.9$ Km & $1.94$ & $92.8\%$ & $0.0\%$ & $65.0\%$ \\ 
TWI & $4.0$ Km & $2.7$ Km & $11.0$ Km & $15.1$ Km & $2.67$ & $92.8\%$ & $25.4\%$ & $34.2\%$ \\ 
DTWBI & $4.0$ Km & $-0.2$ Km & $4.5$ Km & $8.6$ Km & $2.21$ & $95.5\%$ & $10.8\%$ & $25.7\%$ \\ 
DTWBMI-HI & $1.1$ Km & $0.0$ Km & $8.5$ Km & $9.6$ Km & $2.48$ & $94.4\%$ & $17.9\%$ & $27.9\%$ \\ 
DTWBMI-LO & $0.3$ Km & $0.4$ Km & $6.8$ Km & $6.5$ Km & $2.54$ & $94.7\%$ & $18.2\%$ & $22.5\%$ \\ 
\midrule\addlinespace[2.5pt]
\multicolumn{9}{l}{12 hrs} \\ 
\midrule\addlinespace[2.5pt]
LI & $9.4$ Km & $-1.9$ Km & $0.0$ Km & $9.4$ Km & $0.94$ & $94.4\%$ & $0.9\%$ & $56.0\%$ \\ 
MI & $10.9$ Km & $21.2$ Km & $16.3$ Km & $5.3$ Km & $1.18$ & $95.2\%$ & $0.0\%$ & $57.0\%$ \\ 
TWI & $9.3$ Km & $13.0$ Km & $16.9$ Km & $7.6$ Km & $2.13$ & $93.8\%$ & $37.9\%$ & $25.3\%$ \\ 
DTWBI & $0.1$ Km & $-0.4$ Km & $7.0$ Km & $6.9$ Km & $1.41$ & $95.9\%$ & $9.4\%$ & $35.2\%$ \\ 
DTWBMI-HI & $4.5$ Km & $2.4$ Km & $11.3$ Km & $6.8$ Km & $1.63$ & $94.3\%$ & $26.0\%$ & $28.5\%$ \\ 
DTWBMI-LO & $0.2$ Km & $1.7$ Km & $6.1$ Km & $5.9$ Km & $1.44$ & $96.0\%$ & $18.9\%$ & $23.9\%$ \\ 
\bottomrule
\end{longtable}

Table 3 shows performance of the imputation methods across time-based
conditions in the 3-hour gap condition. Because people are less likely
to travel during the night hours, different methodology or parameter
sets may be preferable when imputing missing data occurring only at
night. Records in which the missingness was induced only in the hours
between 22:00 and 05:00 were identified and marked as ``Night Only,'' to
be compared to records also containing missingness occurring during the
daytime hours. Missingness occurring only at night was imputed with less
absolute bias in the DTWBMI-LO, DTWBI, and linear interpolation
conditions. Systematic over- and underestimation was lower across all
methods except for mean imputation when the missingness occurred at
night. DTWBMI-LO and DTWBI offer the best daytime performance, with
comparable absolute bias and median bias. DTWBMI-HI demonstrates less
systematic bias towards underestimation of travel periods than DTWBI.

\begin{longtable}{l|rrrrrrrr}
\caption*{
{\large Table 3: Method comparison across night only vs day} \\ 
{\small 3 hour gap condition}
} \\ 
\toprule
\multicolumn{1}{l}{} & Abs Bias & Med Bias & Dist Over & Dist Under & RMSE & TP Acc. & TP Over & TP Under \\ 
\midrule\addlinespace[2.5pt]
\multicolumn{9}{l}{Daytime Missing} \\ 
\midrule\addlinespace[2.5pt]
LI & $3.2$ Km & $-0.4$ Km & $0.0$ Km & $3.2$ Km & $0.99$ & $89.5\%$ & $3.3\%$ & $36.8\%$ \\ 
MI & $0.3$ Km & $5.0$ Km & $3.7$ Km & $4.0$ Km & $1.27$ & $90.8\%$ & $0.0\%$ & $42.1\%$ \\ 
TWI & $2.3$ Km & $3.4$ Km & $6.9$ Km & $4.6$ Km & $2.17$ & $87.5\%$ & $23.1\%$ & $27.4\%$ \\ 
DTWBI & $1.3$ Km & $0.0$ Km & $1.3$ Km & $2.6$ Km & $1.39$ & $92.8\%$ & $8.2\%$ & $22.0\%$ \\ 
DTWBMI-HI & $2.7$ Km & $0.0$ Km & $1.3$ Km & $4.0$ Km & $1.34$ & $91.5\%$ & $9.5\%$ & $25.9\%$ \\ 
DTWBMI-LO & $1.2$ Km & $0.0$ Km & $1.0$ Km & $2.2$ Km & $1.37$ & $92.8\%$ & $9.8\%$ & $18.3\%$ \\ 
\midrule\addlinespace[2.5pt]
\multicolumn{9}{l}{Night Only} \\ 
\midrule\addlinespace[2.5pt]
LI & $0.0$ Km & $0.0$ Km & $0.0$ Km & $0.0$ Km & $0.01$ & $100.0\%$ & $0.0\%$ & $0.0\%$ \\ 
MI & $5.8$ Km & $5.8$ Km & $5.8$ Km & $0.0$ Km & $0.48$ & $100.0\%$ & $0.0\%$ & $0.0\%$ \\ 
TWI & $0.6$ Km & $0.1$ Km & $0.6$ Km & $0.0$ Km & $0.14$ & $99.4\%$ & $4.6\%$ & $0.0\%$ \\ 
DTWBI & $0.0$ Km & $0.0$ Km & $0.0$ Km & $0.0$ Km & $0.01$ & $100.0\%$ & $0.0\%$ & $0.0\%$ \\ 
DTWBMI-HI & $0.5$ Km & $0.0$ Km & $0.5$ Km & $0.0$ Km & $0.11$ & $99.4\%$ & $4.2\%$ & $0.0\%$ \\ 
DTWBMI-LO & $0.0$ Km & $0.0$ Km & $0.1$ Km & $0.0$ Km & $0.02$ & $99.8\%$ & $0.8\%$ & $0.0\%$ \\ 
\bottomrule
\end{longtable}

As the data contained multiple different sets of data for some
respondents, it was possible to investigate the amount to which access
to someone's own complete travel behavior from different time points
could be beneficial in the imputation procedure. In Table 4, the methods
are compared against the quantity of own data available as different
data sets for serving as imputation candidates. For persons where only a
single set was available and whom therefore could not serve as their own
imputation candidate, DTWBMI-LO was the preferred method on the basis of
all measures. When 2-3 sets were available, DTWBMI-HI and DTWBMI-LO
provided similar performance with respect to the mean bias and over- and
under-estimation. However, the travel period metrics demonstrate
preference for the DTWBMI-LO method. In the 4+ sets condition, MI had
the lowest absolute bias despite otherwise poor performance, while
DTWBMI-LO had a preferable profile for reduction of systematic bias both
for travel periods and distance. Notably, the performance of all DTW
methods worsens as the number of sets increases, running counter to
expectations.

\begin{longtable}{l|rrrrrrrr}
\caption*{
{\large Table 4: Method comparison across number of reference sets} \\ 
{\small Imputing travel distance}
} \\ 
\toprule
\multicolumn{1}{l}{} & Abs Bias & Med Bias & Dist Over & Dist Under & RMSE & TP Acc. & TP Over & TP Under \\ 
\midrule\addlinespace[2.5pt]
\multicolumn{9}{l}{No extra data} \\ 
\midrule\addlinespace[2.5pt]
LI & $4.8$ Km & $-0.6$ Km & $0.0$ Km & $4.8$ Km & $1.01$ & $91.9\%$ & $4.3\%$ & $38.8\%$ \\ 
MI & $3.7$ Km & $5.6$ Km & $7.7$ Km & $4.0$ Km & $1.32$ & $94.0\%$ & $0.0\%$ & $45.3\%$ \\ 
TWI & $3.7$ Km & $2.3$ Km & $9.6$ Km & $5.9$ Km & $2.07$ & $91.4\%$ & $25.2\%$ & $24.9\%$ \\ 
DTWBI & $0.5$ Km & $-0.1$ Km & $3.7$ Km & $4.1$ Km & $1.41$ & $94.7\%$ & $9.2\%$ & $22.1\%$ \\ 
DTWBMI-HI & $0.7$ Km & $0.0$ Km & $5.0$ Km & $4.3$ Km & $1.50$ & $94.1\%$ & $17.1\%$ & $22.7\%$ \\ 
DTWBMI-LO & $0.2$ Km & $0.0$ Km & $3.4$ Km & $3.1$ Km & $1.45$ & $95.0\%$ & $14.6\%$ & $17.1\%$ \\ 
\midrule\addlinespace[2.5pt]
\multicolumn{9}{l}{2-3 sets} \\ 
\midrule\addlinespace[2.5pt]
LI & $6.8$ Km & $-0.4$ Km & $0.0$ Km & $6.8$ Km & $1.09$ & $92.0\%$ & $3.0\%$ & $40.6\%$ \\ 
MI & $1.6$ Km & $5.7$ Km & $7.6$ Km & $6.0$ Km & $1.42$ & $93.0\%$ & $0.0\%$ & $45.5\%$ \\ 
TWI & $1.2$ Km & $2.4$ Km & $8.6$ Km & $7.4$ Km & $2.01$ & $91.6\%$ & $22.9\%$ & $24.5\%$ \\ 
DTWBI & $1.4$ Km & $0.0$ Km & $3.2$ Km & $4.6$ Km & $1.45$ & $95.5\%$ & $7.5\%$ & $21.9\%$ \\ 
DTWBMI-HI & $0.4$ Km & $0.0$ Km & $4.8$ Km & $5.2$ Km & $1.51$ & $94.0\%$ & $12.0\%$ & $21.7\%$ \\ 
DTWBMI-LO & $0.4$ Km & $0.0$ Km & $3.6$ Km & $4.0$ Km & $1.44$ & $95.3\%$ & $11.9\%$ & $16.1\%$ \\ 
\midrule\addlinespace[2.5pt]
\multicolumn{9}{l}{4+ sets} \\ 
\midrule\addlinespace[2.5pt]
LI & $5.6$ Km & $-0.1$ Km & $0.0$ Km & $5.6$ Km & $1.02$ & $91.9\%$ & $4.2\%$ & $28.5\%$ \\ 
MI & $0.8$ Km & $4.0$ Km & $8.1$ Km & $7.3$ Km & $1.36$ & $94.5\%$ & $0.0\%$ & $34.3\%$ \\ 
TWI & $1.6$ Km & $1.5$ Km & $7.0$ Km & $8.6$ Km & $1.80$ & $92.8\%$ & $22.6\%$ & $19.6\%$ \\ 
DTWBI & $3.7$ Km & $0.0$ Km & $1.0$ Km & $4.7$ Km & $1.32$ & $96.2\%$ & $5.5\%$ & $17.6\%$ \\ 
DTWBMI-HI & $2.4$ Km & $0.0$ Km & $4.7$ Km & $7.1$ Km & $1.48$ & $94.6\%$ & $11.9\%$ & $19.1\%$ \\ 
DTWBMI-LO & $1.7$ Km & $0.0$ Km & $2.5$ Km & $4.2$ Km & $1.43$ & $95.6\%$ & $10.3\%$ & $15.2\%$ \\ 
\bottomrule
\end{longtable}

To determine whether this result was due to inherent differences between
groups with varying amounts of extra data, for example due to a
relationship between more travel behavior leading to longer tracking
times, an additional simulation study was conducted using only those
persons with at least four sets. The details are described in
\protect\hyperlink{sec:appendixb}{Appendix B}. In this restricted
simulation study, performance does indeed improve with additional own
data sets, but the preference for the high-information DTWBMI-HI is not
demonstrated.

\hypertarget{number-of-trips}{%
\subsection{Number of trips}\label{number-of-trips}}

DTWB(M)I methods are not restricted to the imputation of travel
distance. All travel metrics that occur as a function of time may
benefit from imputation as time series. Table 5 presents a comparison of
the same methods applied to the number of trips occurring within a gap.
While all methods present with very little bias, the methods DTWBMI-LO
and DTWBI offer an improved travel period accuracy and less systematic
overestimation of the number of periods containing movement.

\begin{longtable}{l|rrrrrr}
\caption*{
{\large Table 5: Method comparison across all cases} \\ 
{\small Imputing number of trips}
} \\ 
\toprule
\multicolumn{1}{l}{} & Abs Bias & Med Bias & RMSE & TP Over & TP Under & TP Acc. \\ 
\midrule\addlinespace[2.5pt]
LI & $0.02$ & $0.00$ & $0.30$ & $23.8\%$ & $10.0\%$ & $74.9\%$ \\ 
TWI & $0.04$ & $0.00$ & $0.27$ & $26.1\%$ & $20.0\%$ & $91.2\%$ \\ 
DTWBI & $0.03$ & $0.00$ & $0.21$ & $14.0\%$ & $15.0\%$ & $94.4\%$ \\ 
DTWBMI-HI & $0.02$ & $0.00$ & $0.21$ & $12.7\%$ & $19.4\%$ & $94.3\%$ \\ 
DTWBMI-LO & $0.02$ & $0.00$ & $0.21$ & $14.4\%$ & $14.4\%$ & $94.7\%$ \\ 
\bottomrule
\end{longtable}

Table 6 shows the results of the simulations stratified by the length of
missingness. While linear interpolation performs with more bias as the
length of the missingness increases, the DTWBMI-LO and DTWBI methods
deliver relatively consistent performance with regards to average
absolute bias, and offer better relative performance in travel period
accuracy and over- and underestimation. At a gap length of 12 hours, all
DTW-based methods maintain a travel period accuracy of approximately
95\%, while linear interpolation offers only a 63\% travel period
accuracy.

\begin{longtable}{l|rrrrrrr}
\caption*{
{\large Table 6: Method comparison across varying gap lengths} \\ 
{\small Imputing number of trips}
} \\ 
\toprule
\multicolumn{1}{l}{} & Abs Bias & Trips & Med Bias & RMSE & TP Over & TP Under & TP Acc. \\ 
\midrule\addlinespace[2.5pt]
\multicolumn{8}{l}{1 hr} \\ 
\midrule\addlinespace[2.5pt]
LI & $0.000$ & 1.03 & $0.000$ & $0.10$ & $6.5\%$ & $1.0\%$ & $93.2\%$ \\ 
TWI & $0.002$ & 1.03 & $0.000$ & $0.21$ & $15.3\%$ & $11.1\%$ & $87.2\%$ \\ 
DTWBI & $0.000$ & 1.03 & $0.000$ & $0.13$ & $4.5\%$ & $9.8\%$ & $92.5\%$ \\ 
DTWBMI-HI & $0.000$ & 1.03 & $0.000$ & $0.09$ & $0.9\%$ & $8.4\%$ & $95.3\%$ \\ 
DTWBMI-LO & $0.003$ & 1.03 & $0.000$ & $0.11$ & $4.7\%$ & $4.5\%$ & $95.1\%$ \\ 
\midrule\addlinespace[2.5pt]
\multicolumn{8}{l}{3 hrs} \\ 
\midrule\addlinespace[2.5pt]
LI & $0.000$ & 0.98 & $0.000$ & $0.17$ & $14.2\%$ & $3.0\%$ & $85.2\%$ \\ 
TWI & $0.025$ & 0.98 & $0.000$ & $0.21$ & $23.5\%$ & $12.5\%$ & $91.5\%$ \\ 
DTWBI & $0.000$ & 0.98 & $0.000$ & $0.14$ & $9.2\%$ & $8.7\%$ & $95.7\%$ \\ 
DTWBMI-HI & $0.007$ & 0.98 & $0.000$ & $0.13$ & $6.7\%$ & $11.5\%$ & $95.5\%$ \\ 
DTWBMI-LO & $0.012$ & 0.98 & $0.000$ & $0.13$ & $7.7\%$ & $8.4\%$ & $96.2\%$ \\ 
\midrule\addlinespace[2.5pt]
\multicolumn{8}{l}{6 hrs} \\ 
\midrule\addlinespace[2.5pt]
LI & $0.000$ & 0.95 & $0.000$ & $0.32$ & $26.0\%$ & $7.0\%$ & $72.8\%$ \\ 
TWI & $0.030$ & 0.95 & $0.000$ & $0.28$ & $25.7\%$ & $21.9\%$ & $91.6\%$ \\ 
DTWBI & $0.010$ & 0.95 & $0.000$ & $0.23$ & $12.0\%$ & $15.8\%$ & $94.2\%$ \\ 
DTWBMI-HI & $0.013$ & 0.95 & $0.000$ & $0.24$ & $15.9\%$ & $20.0\%$ & $92.9\%$ \\ 
DTWBMI-LO & $0.015$ & 0.95 & $0.000$ & $0.24$ & $14.3\%$ & $14.6\%$ & $94.1\%$ \\ 
\midrule\addlinespace[2.5pt]
\multicolumn{8}{l}{10 hrs} \\ 
\midrule\addlinespace[2.5pt]
LI & $0.040$ & 0.98 & $0.000$ & $0.46$ & $37.2\%$ & $16.0\%$ & $60.5\%$ \\ 
TWI & $0.029$ & 0.98 & $0.000$ & $0.32$ & $33.4\%$ & $26.7\%$ & $92.5\%$ \\ 
DTWBI & $0.030$ & 0.98 & $0.000$ & $0.29$ & $25.4\%$ & $17.6\%$ & $94.4\%$ \\ 
DTWBMI-HI & $0.003$ & 0.98 & $0.000$ & $0.28$ & $20.2\%$ & $25.8\%$ & $93.5\%$ \\ 
DTWBMI-LO & $0.007$ & 0.98 & $0.000$ & $0.29$ & $24.6\%$ & $19.3\%$ & $93.7\%$ \\ 
\midrule\addlinespace[2.5pt]
\multicolumn{8}{l}{12 hrs} \\ 
\midrule\addlinespace[2.5pt]
LI & $0.060$ & 0.98 & $0.000$ & $0.44$ & $35.1\%$ & $23.0\%$ & $62.6\%$ \\ 
TWI & $0.027$ & 0.98 & $0.000$ & $0.32$ & $32.6\%$ & $27.6\%$ & $93.2\%$ \\ 
DTWBI & $0.030$ & 0.98 & $0.000$ & $0.27$ & $18.7\%$ & $23.0\%$ & $95.4\%$ \\ 
DTWBMI-HI & $0.003$ & 0.98 & $0.000$ & $0.28$ & $20.0\%$ & $31.5\%$ & $94.2\%$ \\ 
DTWBMI-LO & $0.000$ & 0.98 & $0.000$ & $0.28$ & $20.7\%$ & $25.2\%$ & $94.5\%$ \\ 
\bottomrule
\end{longtable}

\hypertarget{sec:conclusion}{%
\section{Conclusion}\label{sec:conclusion}}

In this research, we introduced a unique approach to imputing travel
behavior characteristics in human trajectory data, which we call Dynamic
Time Warping Based Multiple Imputation (DTWBMI). We tested the
performance of this methodology on a real-life dataset and conducted a
simulation study to gauge the impact of model parameters. The outcomes
convincingly indicated that the DTWBMI technique superseded other
gap-filling strategies, such as mean imputation, Time Window Imputation,
and linear interpolation, specifically for long gaps.

We demonstrate that different methods for gap-filling may provide better
or worse results on the basis of the nature of the missingness,
e.g.~depending on gap length or time or day, or the nature of the data
itself, e.g.~depending on the number of own reference sets. Linear
interpolation is an appropriate method for small gaps, and when
missingness occurs at night, because the chance of travel is generally
low in these situations. On the other hand, in situations where it is
necessary to fill a long gap, imputation methods will likely provide
superior results because of their capacity to appropriately consider
travel behavior variance. When a sufficient quantity of a person's own
data is available to use as donor candidates, methods selecting fewer
candidates with more closely aligned travel behavior should be preferred
for their capacity to reduce the variance of the estimate.

DTWBMI-HI, with an 8-hour matching buffer and high candidate specificity
was expected to be able to match on the basis of longer travel behavior
patterns, such as commuting behavior, and was expected to have an
advantage over DTWBMI-LO in scenarios in which travel behavior was more
predictable, such as in the Daytime Missing or 4+ reference set
conditions. In fact, DTWBMI-HI underperformed relative to DTWBMI-LO in
almost all scenarios, including when imputation was restricted to
persons with additional own data sets as in
\protect\hyperlink{sec:appendixb}{Appendix B}. DTWBMI-LO is defined by a
short 1-hour matching buffer, medium candidate specificity, and
unrestricted time window, and thus matched imputation candidates on the
travel behavior immediately preceding and following the gap. This proved
to be a good fit across many scenarios, including as the length of the
gap increased to 10 and even 12 hours.

Comparisons between the performance of DTWBMI-LO and DTWBI across
scenarios contextualize the benefit of single versus multiple
imputation. While DTWBI sometimes had the lowest absolute bias, the
method tended toward a larger systematic underestimation. DTWBMI-LO
performed more consistently across scenarios, despite its systematic
overestimation of distance and travel periods. The source of this
systematic bias is unclear, but may be due to differences in imputation
candidates available under the two parameter sets. The longer match
buffer and higher candidate specificity of DTWBMI-HI may more often be
restricted to only the longer sets, where there is a slight negative
correlation with distance traveled per period. DTWBMI-LO imputes on
average an equal number of travel period overestimations and
underestimations, indicating that the distance per period is too high.

We recognize certain constraints associated with the DTWBMI method,
including the necessity for at least one complete dataset and the
requirement for extended observation periods. Consequently, we recommend
future research endeavors focus on investigating the limits of
feasibility of the DTWBMI method for imputing travel behavior, and in
broadening its potential applications.

In terms of future directions, we suggest examination of optimal
parameters in datasets encompassing a larger population, prolonged
participation periods, or data exhibiting Missing Not At Random
characteristics. This will allow for establishing the generalizability
and robustness of the DTWBMI method, and may additionally shed light on
the appropriateness of varying parameter sets for differing datasets.
Given that travel mode is a fundamental component of travel diary
studies, it would also be worth exploring the possibility of imputing
travel mode information concurrently with travel behavior
characteristics via the DTWBMI method.

Notwithstanding these limitations, we posit that the DTWBMI method holds
pragmatic implications for augmenting the precision of app-based travel
diaries, and addressing its pervasive missing data problems,
particularly when external data resources are limited. We urge
researchers and practitioners to recognize this methodology as a
potential solution for filling long gaps in human mobility data and to
consider its capacity for expansion in various contexts.

\hypertarget{references}{%
\section*{References}\label{references}}
\addcontentsline{toc}{section}{References}

\hypertarget{refs}{}
\begin{CSLReferences}{1}{0}
\leavevmode\vadjust pre{\hypertarget{ref-Bahr2020-mt}{}}%
Bähr, S., Haas, G.-C., Keusch, F., Kreuter, F., Trappmann, M., 2020.
Missing data and other measurement quality issues in mobile geolocation
sensor data. Social science computer review 0894439320944118.
doi:\href{https://doi.org/10.1177/0894439320944118}{10.1177/0894439320944118}

\leavevmode\vadjust pre{\hypertarget{ref-Barnett2020-rr}{}}%
Barnett, I., Onnela, J.-P., 2020. Inferring mobility measures from {GPS}
traces with missing data. Biostatistics 21, e98--e112.
doi:\href{https://doi.org/10.1093/biostatistics/kxy059}{10.1093/biostatistics/kxy059}

\leavevmode\vadjust pre{\hypertarget{ref-Beukenhorst2022-rk}{}}%
Beukenhorst, A.L., Druce, K.L., De Cock, D., 2022. Smartphones for
musculoskeletal research--hype or hope? Lessons from a decennium of
mHealth studies. BMC Musculoskeletal Disorders 23, 487.

\leavevmode\vadjust pre{\hypertarget{ref-Beukenhorst2021-ci}{}}%
Beukenhorst, A.L., Sergeant, J.C., Schultz, D.M., McBeth, J., Yimer,
B.B., Dixon, W.G., 2021. Understanding the predictors of missing
location data to inform smartphone study design: Observational study.
JMIR mHealth and uHealth 9, e28857.
doi:\href{https://doi.org/10.2196/28857}{10.2196/28857}

\leavevmode\vadjust pre{\hypertarget{ref-boeschoten2020digital}{}}%
Boeschoten, L., Ausloos, J., Moeller, J., Araujo, T., Oberski, D.L.,
2020. Digital trace data collection through data donation. arXiv
preprint arXiv:2011.09851.

\leavevmode\vadjust pre{\hypertarget{ref-Van_Buuren2018-hw}{}}%
Buuren, S. van, 2018. Flexible imputation of missing data, second
edition. CRC Press.

\leavevmode\vadjust pre{\hypertarget{ref-Centraal_Bureau_voor_de_Statistiek2022-fk}{}}%
Centraal Bureau voor de Statistiek, 2022.
\href{https://www.cbs.nl/nl-nl/longread/rapportages/2022/onderweg-in-nederland--odin---2018-2020}{Onderweg
in nederland ({ODiN}) 2018-2020}.

\leavevmode\vadjust pre{\hypertarget{ref-Chen2019-cn}{}}%
Chen, G., Viana, A.C., Fiore, M., Sarraute, C., 2019. Complete
trajectory reconstruction from sparse mobile phone data. EPJ Data
Science 8, 30.
doi:\href{https://doi.org/10.1140/epjds/s13688-019-0206-8}{10.1140/epjds/s13688-019-0206-8}

\leavevmode\vadjust pre{\hypertarget{ref-Dekker2022-wz}{}}%
Dekker, L., elić, K., Dijk, M. van, Holtrop, S., Keuper, N., Laan, L.
van der, Leeuwen, T. van, Meijs, C., Schroten, H., Wijk, L. van, 2022.
Interpolating location data with brownian motion. arXiv preprint
arXiv:2207.01618.

\leavevmode\vadjust pre{\hypertarget{ref-Dhont2021-ab}{}}%
Dhont, M., Tsiporkova, E., González-Deleito, N., 2021. Deriving
spatio-temporal trajectory fingerprints from mobility data using
{Non-Negative} matrix factorisation, in: 2021 International Conference
on Data Mining Workshops ({ICDMW}). ieeexplore.ieee.org, pp. 750--759.
doi:\href{https://doi.org/10.1109/ICDMW53433.2021.00098}{10.1109/ICDMW53433.2021.00098}

\leavevmode\vadjust pre{\hypertarget{ref-Furletti2013-bs}{}}%
Furletti, B., Cintia, P., Renso, C., Spinsanti, L., 2013. Inferring
human activities from GPS tracks, in: Proceedings of the 2nd ACM SIGKDD
International Workshop on Urban Computing. pp. 1--8.

\leavevmode\vadjust pre{\hypertarget{ref-Gonzalez-Perez2022-uq}{}}%
González-Pérez, A., Matey-Sanz, M., Granell, C., Casteleyn, S., 2022.
Using mobile devices as scientific measurement instruments: Reliable
android task scheduling. Pervasive and Mobile Computing 81, 101550.

\leavevmode\vadjust pre{\hypertarget{ref-Harding2021-aq}{}}%
Harding, C., Faghih Imani, A., Srikukenthiran, S., Miller, E.J., others,
2021. Are we there yet? Assessing smartphone apps as full-fledged tools
for activity-travel surveys. Transportation.

\leavevmode\vadjust pre{\hypertarget{ref-Hawthorne2005-zs}{}}%
Hawthorne, G., Elliott, P., 2005. Imputing cross-sectional missing data:
Comparison of common techniques. The Australian and New Zealand journal
of psychiatry 39, 583--590.
doi:\href{https://doi.org/10.1080/j.1440-1614.2005.01630.x}{10.1080/j.1440-1614.2005.01630.x}

\leavevmode\vadjust pre{\hypertarget{ref-hollingshead2021ethics}{}}%
Hollingshead, W., Quan-Haase, A., Chen, W., 2021. Ethics and privacy in
computational social science: A call for pedagogy, in: Handbook of
Computational Social Science, Volume 1. Routledge, pp. 171--185.

\leavevmode\vadjust pre{\hypertarget{ref-Honaker2010-pl}{}}%
Honaker, J., King, G., 2010. What to do about missing values in
time-series cross-section data. American journal of political science
54, 561--581.
doi:\href{https://doi.org/10.1111/j.1540-5907.2010.00447.x}{10.1111/j.1540-5907.2010.00447.x}

\leavevmode\vadjust pre{\hypertarget{ref-Hung2022-js}{}}%
Hung, L.-C., Hu, Y.-H., Tsai, C.-F., Huang, M.-W., 2022. A dynamic time
warping approach for handling class imbalanced medical datasets with
missing values: A case study of protein localization site prediction.
Expert systems with applications 192, 116437.
doi:\href{https://doi.org/10.1016/j.eswa.2021.116437}{10.1016/j.eswa.2021.116437}

\leavevmode\vadjust pre{\hypertarget{ref-Jagadeesh2017-yq}{}}%
Jagadeesh, G.R., Srikanthan, T., 2017. Online {Map-Matching} of noisy
and sparse location data with hidden markov and route choice models.
IEEE Transactions on Intelligent Transportation Systems 18, 2423--2434.
doi:\href{https://doi.org/10.1109/TITS.2017.2647967}{10.1109/TITS.2017.2647967}

\leavevmode\vadjust pre{\hypertarget{ref-Karaim2018-mp}{}}%
Karaim, M., Elsheikh, M., Noureldin, A., Rustamov, R.B., 2018. {GNSS}
error sources. Multifunctional Operation and Application of GPS 69--85.

\leavevmode\vadjust pre{\hypertarget{ref-keusch2022using}{}}%
Keusch, F., Conrad, F.G., 2022. Using smartphones to capture and combine
self-reports and passively measured behavior in social research. Journal
of Survey Statistics and Methodology 10, 863--885.

\leavevmode\vadjust pre{\hypertarget{ref-Keusch2022-cm}{}}%
Keusch, F., Wenz, A., Conrad, F., 2022. Do you have your smartphone with
you? Behavioral barriers for measuring everyday activities with
smartphone sensors. Computers in human behavior 127.
doi:\href{https://doi.org/10.1016/j.chb.2021.107054}{10.1016/j.chb.2021.107054}

\leavevmode\vadjust pre{\hypertarget{ref-Knapen2018-vd}{}}%
Knapen, L., Bellemans, T., Janssens, D., Wets, G., 2018.
Likelihood-based offline map matching of {GPS} recordings using global
trace information. Transportation Research Part C: Emerging Technologies
93, 13--35.
doi:\href{https://doi.org/10.1016/j.trc.2018.05.014}{10.1016/j.trc.2018.05.014}

\leavevmode\vadjust pre{\hypertarget{ref-Kostadinova2012-ex}{}}%
Kostadinova, E., Boeva, V., Boneva, L., Tsiporkova, E., 2012. An
integrative {DTW-based} imputation method for gene expression time
series data, in: 2012 6th {IEEE} International Conference Intelligent
Systems. pp. 258--263.
doi:\href{https://doi.org/10.1109/IS.2012.6335145}{10.1109/IS.2012.6335145}

\leavevmode\vadjust pre{\hypertarget{ref-Kreuter2020-gj}{}}%
Kreuter, F., Haas, G.-C., Keusch, F., Bähr, S., Trappmann, M., 2020.
Collecting survey and smartphone sensor data with an app: Opportunities
and challenges around privacy and informed consent. Social science
computer review 38, 533--549.
doi:\href{https://doi.org/10.1177/0894439318816389}{10.1177/0894439318816389}

\leavevmode\vadjust pre{\hypertarget{ref-Little1988-aq}{}}%
Little, R.J.A., 1988. {Missing-Data} adjustments in large surveys.
Journal of business \& economic statistics: a publication of the
American Statistical Association 6, 287--296.
doi:\href{https://doi.org/10.1080/07350015.1988.10509663}{10.1080/07350015.1988.10509663}

\leavevmode\vadjust pre{\hypertarget{ref-McCool2021-ae}{}}%
McCool, D., Lugtig, P., Mussmann, O., Schouten, B., 2021. An
app-assisted travel survey in official statistics: Possibilities and
challenges. Journal of official statistics 37, 149--170.
doi:\href{https://doi.org/10.2478/jos-2021-0007}{10.2478/jos-2021-0007}

\leavevmode\vadjust pre{\hypertarget{ref-mccool2022maximum}{}}%
McCool, D., Lugtig, P., Schouten, B., 2022. Maximum interpolable gap
length in missing smartphone-based GPS mobility data. Transportation
1--31.

\leavevmode\vadjust pre{\hypertarget{ref-Mennis2018-vz}{}}%
Mennis, J., Mason, M., Coffman, D.L., Henry, K., 2018. Geographic
imputation of missing activity space data from ecological momentary
assessment ({EMA}) {GPS} positions. International journal of
environmental research and public health 15.
doi:\href{https://doi.org/10.3390/ijerph15122740}{10.3390/ijerph15122740}

\leavevmode\vadjust pre{\hypertarget{ref-Meratnia2004-mv}{}}%
Meratnia, N., By, R.A. de, 2004. Spatiotemporal compression techniques
for moving point objects, in: Advances in Database Technology - {EDBT}
2004. Springer Berlin Heidelberg, pp. 765--782.
doi:\href{https://doi.org/10.1007/978-3-540-24741-8/_44}{10.1007/978-3-540-24741-8\textbackslash\_44}

\leavevmode\vadjust pre{\hypertarget{ref-Moffat2007-pj}{}}%
Moffat, A.M., Papale, D., Reichstein, M., Hollinger, D.Y., Richardson,
A.D., Barr, A.G., Beckstein, C., Braswell, B.H., Churkina, G., Desai,
A.R., Falge, E., Gove, J.H., Heimann, M., Hui, D., Jarvis, A.J., Kattge,
J., Noormets, A., Stauch, V.J., 2007. Comprehensive comparison of
gap-filling techniques for eddy covariance net carbon fluxes.
Agricultural and Forest Meteorology 147, 209--232.
doi:\href{https://doi.org/10.1016/j.agrformet.2007.08.011}{10.1016/j.agrformet.2007.08.011}

\leavevmode\vadjust pre{\hypertarget{ref-montoliu2013discovering}{}}%
Montoliu, R., Blom, J., Gatica-Perez, D., 2013. Discovering places of
interest in everyday life from smartphone data. Multimedia tools and
applications 62, 179--207.

\leavevmode\vadjust pre{\hypertarget{ref-Onnela2021-is}{}}%
Onnela, J.-P., 2021. Opportunities and challenges in the collection and
analysis of digital phenotyping data. Neuropsychopharmacology: official
publication of the American College of Neuropsychopharmacology 46,
45--54.
doi:\href{https://doi.org/10.1038/s41386-020-0771-3}{10.1038/s41386-020-0771-3}

\leavevmode\vadjust pre{\hypertarget{ref-Park2022-jn}{}}%
Park, J., Muller, J., Arora, B., Faybishenko, B., Pastorello, G.,
Varadharajan, C., Sahu, R., Agarwal, D., 2022.
\href{https://arxiv.org/abs/2202.12441}{{Long-Term} missing value
imputation for time series data using deep neural networks}.

\leavevmode\vadjust pre{\hypertarget{ref-Parrella2021-gn}{}}%
Parrella, M.L., Albano, G., Perna, C., La Rocca, M., 2021. Bootstrap
joint prediction regions for sequences of missing values in
spatio-temporal datasets. Computational statistics 36, 2917--2938.
doi:\href{https://doi.org/10.1007/s00180-021-01099-y}{10.1007/s00180-021-01099-y}

\leavevmode\vadjust pre{\hypertarget{ref-Phan2018-fu}{}}%
Phan, T.-T.-H., Bigand, A., Caillault, É.P., 2018a. A new fuzzy
logic-based similarity measure applied to large gap imputation for
uncorrelated multivariate time series. Applied computational
intelligence and soft computing 2018, 1--15.
doi:\href{https://doi.org/10.1155/2018/9095683}{10.1155/2018/9095683}

\leavevmode\vadjust pre{\hypertarget{ref-Phan2018-tf}{}}%
Phan, T.-T.-H., Caillault, É.P., Bigand, A., 2018b. Comparative study on
univariate forecasting methods for meteorological time series, in: 2018
26th European Signal Processing Conference ({EUSIPCO}).
ieeexplore.ieee.org, pp. 2380--2384.
doi:\href{https://doi.org/10.23919/EUSIPCO.2018.8553576}{10.23919/EUSIPCO.2018.8553576}

\leavevmode\vadjust pre{\hypertarget{ref-Phan2020-qg}{}}%
Phan, T.-T.-H., Poisson Caillault, É., Bigand, A., 2020a. {eDTWBI}:
Effective imputation method for univariate time series, in: Advanced
Computational Methods for Knowledge Engineering. Springer International
Publishing, pp. 121--132.
doi:\href{https://doi.org/10.1007/978-3-030-38364-0/_11}{10.1007/978-3-030-38364-0\textbackslash\_11}

\leavevmode\vadjust pre{\hypertarget{ref-Phan2020-kc}{}}%
Phan, T.-T.-H., Poisson Caillault, É., Lefebvre, A., Bigand, A., 2020b.
Dynamic time warping-based imputation for univariate time series data.
Pattern recognition letters 139, 139--147.
doi:\href{https://doi.org/10.1016/j.patrec.2017.08.019}{10.1016/j.patrec.2017.08.019}

\leavevmode\vadjust pre{\hypertarget{ref-Rabiner1993-vp}{}}%
Rabiner, L., Juang, B.-H., 1993. Fundamentals of speech recognition.
Prentice-Hall, Inc., USA.

\leavevmode\vadjust pre{\hypertarget{ref-Ranasinghe2018-ky}{}}%
Ranasinghe, C., Kray, C., 2018. Location information quality: A review.
Sensors 18.
doi:\href{https://doi.org/10.3390/s18113999}{10.3390/s18113999}

\leavevmode\vadjust pre{\hypertarget{ref-Rout2021-tj}{}}%
Rout, A., Nitoslawski, S., Ladle, A., Galpern, P., 2021. Using
{smartphone-GPS} data to understand pedestrian-scale behavior in urban
settings: A review of themes and approaches. Computers, environment and
urban systems 90, 101705.
doi:\href{https://doi.org/10.1016/j.compenvurbsys.2021.101705}{10.1016/j.compenvurbsys.2021.101705}

\leavevmode\vadjust pre{\hypertarget{ref-Rubin2004-tg}{}}%
Rubin, D.B., 2004. Multiple imputation for nonresponse in surveys. John
Wiley \& Sons.

\leavevmode\vadjust pre{\hypertarget{ref-Sakoe1978-pl}{}}%
Sakoe, H., Chiba, S., 1978. Dynamic programming algorithm optimization
for spoken word recognition. IEEE transactions on acoustics, speech, and
signal processing 26, 43--49.
doi:\href{https://doi.org/10.1109/TASSP.1978.1163055}{10.1109/TASSP.1978.1163055}

\leavevmode\vadjust pre{\hypertarget{ref-Siddique2008-yh}{}}%
Siddique, J., Belin, T.R., 2008. Multiple imputation using an iterative
hot-deck with distance-based donor selection. Statistics in medicine 27,
83--102. doi:\href{https://doi.org/10.1002/sim.3001}{10.1002/sim.3001}

\leavevmode\vadjust pre{\hypertarget{ref-silber2021linking}{}}%
Silber, H., Keusch, F., Breuer, J., Siegers, P., Beuthner, C., Stier,
S., Gummer, T., Weiß, B., 2021. Linking surveys and digital trace data:
Insights from two studies on determinants of data sharing behavior.
SocArXiv Papers.

\leavevmode\vadjust pre{\hypertarget{ref-Stanley2018-rx}{}}%
Stanley, K., Yoo, E.-H., Paul, T., Bell, S., 2018. How many days are
enough?: Capturing routine human mobility. International journal of
geographical information science: IJGIS 32, 1485--1504.
doi:\href{https://doi.org/10.1080/13658816.2018.1434888}{10.1080/13658816.2018.1434888}

\leavevmode\vadjust pre{\hypertarget{ref-stephens2018strategies}{}}%
Stephens, S., Beyene, J., Tremblay, M.S., Faulkner, G., Pullnayegum, E.,
Feldman, B.M., 2018. Strategies for dealing with missing accelerometer
data. Rheumatic Disease Clinics 44, 317--326.

\leavevmode\vadjust pre{\hypertarget{ref-Sun2017-bf}{}}%
Sun, B., Ma, L., Cheng, W., Wen, W., Goswami, P., Bai, G., 2017. An
improved k-nearest neighbours method for traffic time series imputation,
in: 2017 Chinese Automation Congress ({CAC}). ieeexplore.ieee.org, pp.
7346--7351.
doi:\href{https://doi.org/10.1109/CAC.2017.8244105}{10.1109/CAC.2017.8244105}

\leavevmode\vadjust pre{\hypertarget{ref-Tanaka2021-en}{}}%
Tanaka, A., Tateiwa, N., Hata, N., Yoshida, A., Wakamatsu, T., Osafune,
S., Fujisawa, K., 2021. Offline map matching using time-expanded graph
for low-frequency data. Transportation Research Part C: Emerging
Technologies 130, 103265.
doi:\href{https://doi.org/10.1016/j.trc.2021.103265}{10.1016/j.trc.2021.103265}

\leavevmode\vadjust pre{\hypertarget{ref-Tormene2009-he}{}}%
Tormene, P., Giorgino, T., Quaglini, S., Stefanelli, M., 2009. Matching
incomplete time series with dynamic time warping: An algorithm and an
application to post-stroke rehabilitation. Artificial intelligence in
medicine 45, 11--34.
doi:\href{https://doi.org/10.1016/j.artmed.2008.11.007}{10.1016/j.artmed.2008.11.007}

\leavevmode\vadjust pre{\hypertarget{ref-Van_Buuren2000-no}{}}%
Van Buuren, S., Oudshoorn, C.G.M., 2000. Multivariate imputation by
chained equations.

\leavevmode\vadjust pre{\hypertarget{ref-Wojtusiak2021-hw}{}}%
Wojtusiak, J., Nia, R.M., 2021. Location prediction using GPS trackers:
Can machine learning help locate the missing people with dementia?
Internet of Things 13, 100035.

\leavevmode\vadjust pre{\hypertarget{ref-Yoo2020-yy}{}}%
Yoo, E., Roberts, J.E., Eum, Y., Shi, Y., 2020. Quality of hybrid
location data drawn from {GPS‐enabled} mobile phones: Does it matter?
Transactions in GIS 90, 187.
doi:\href{https://doi.org/10.1111/tgis.12612}{10.1111/tgis.12612}

\leavevmode\vadjust pre{\hypertarget{ref-Zhang2019-yt}{}}%
Zhang, J., Mu, X., Fang, J., Yang, Y., 2019. Time series imputation via
integration of revealed information based on the residual shortcut
connection. IEEE Access 7, 102397--102405.
doi:\href{https://doi.org/10.1109/ACCESS.2019.2928641}{10.1109/ACCESS.2019.2928641}

\leavevmode\vadjust pre{\hypertarget{ref-Zheng2008-nk}{}}%
Zheng, Y., Li, Q., Chen, Y., Xie, X., Ma, W.-Y., 2008. Understanding
mobility based on GPS data, in: Proceedings of the 10th International
Conference on Ubiquitous Computing. pp. 312--321.

\leavevmode\vadjust pre{\hypertarget{ref-Zhu2022-nq}{}}%
Zhu, L., Boissy, P., Duval, C., Zou, G., Jog, M., Montero-Odasso, M.,
Speechley, M., 2022. How long should {GPS} recording lengths be to
capture the community mobility of an older clinical population? A
parkinson's example. Sensors 22.
doi:\href{https://doi.org/10.3390/s22020563}{10.3390/s22020563}

\end{CSLReferences}

\newpage{}

\hypertarget{appendix-appendix}{%
\section*{(APPENDIX) Appendix}\label{appendix-appendix}}
\addcontentsline{toc}{section}{(APPENDIX) Appendix}

\hypertarget{sec:appendix}{%
\section*{Appendix A - Parameter selection}\label{sec:appendix}}
\addcontentsline{toc}{section}{Appendix A - Parameter selection}

The initial simulation study was performed in order to determine the
appropriate parameters described in the previous section. We selected
the following two key travel metrics for evaluation of performance:
total distance and number of stops. We compare the parameters on the
basis of Root Mean Square Error (RMSE) and mean absolute bias (Bias), as
well as on a set of metrics developed to assess the accuracy and
directional bias of the imputed travel distance and number of periods
spent moving. RMSE and Bias both assess the accuracy of the underlying
imputed metric in absolute terms. Because different parameter sets
generated either a significant upward bias or downward bias on total
distance, we compared under- and overestimation separately.

Travel Periods Overestimated (TP over) reflects the percentage of
15-minute travel periods imputed that did not exist in the true data
set. Conversely, Travel Periods Underestimated (TP Under) reflects the
percentage of the true number of periods that were spent in movement
that were not reflected in the imputation. Travel Period Accuracy (TP
Acc.) reflects the percentage agreement with the total count of
moving/stationary periods between the true data and the imputed data.

Distance overestimated (Dist Over) reflects only the upward bias of the
imputed values relative to the true distance. Distance Underestimated
(Dist Under) similarly reflects only the downward bias of the imputed
values relative to the true distance. Both are expressed in kilometers.

Table 7 shows the mean performance of each parameter option across the
simulations for distance. Overall, a match buffer of 1 hour is preferred
across most metrics and provides the smallest bias in distance. A high
candidate specificity is generally preferred overall, although the
overall bias is larger. Time window has less of a clear pattern when
examining the mean performance. Most metrics prefer a higher number of
imputations, although the relationship between the average distance bias
and number of imputations prefers a smaller number of imputations.

\begin{longtable}{l|rrrrrrr}
\caption*{
{\large Table 7: Comparison among possible DTWBMI parameter values}
} \\ 
\toprule
\multicolumn{1}{l}{} & Bias (km) & Dist over & Dist under & RMSE & TP acc. & TP over & TP under \\ 
\midrule\addlinespace[2.5pt]
\multicolumn{8}{l}{Candidate Specificity} \\ 
\midrule\addlinespace[2.5pt]
Low & $1.9$ & $4.1$ & $5.7$ & $1.317$ & $95\%$ & $16\%$ & $18\%$ \\ 
Medium & $1.7$ & \cellcolor[HTML]{EEEEEE}{$3.9$} & $5.4$ & \cellcolor[HTML]{EEEEEE}{$1.315$} & $95\%$ & \cellcolor[HTML]{EEEEEE}{$15\%$} & $17\%$ \\ 
High & \cellcolor[HTML]{EEEEEE}{$1.6$} & $4.0$ & \cellcolor[HTML]{EEEEEE}{$5.2$} & $1.327$ & \cellcolor[HTML]{EEEEEE}{$95\%$} & $15\%$ & \cellcolor[HTML]{EEEEEE}{$17\%$} \\ 
\midrule\addlinespace[2.5pt]
\multicolumn{8}{l}{Match Buffer} \\ 
\midrule\addlinespace[2.5pt]
1 hour & \cellcolor[HTML]{EEEEEE}{$1.0$} & \cellcolor[HTML]{EEEEEE}{$3.9$} & \cellcolor[HTML]{EEEEEE}{$4.2$} & $1.317$ & $95\%$ & $16\%$ & \cellcolor[HTML]{EEEEEE}{$14\%$} \\ 
4 hours & $1.6$ & $4.0$ & $5.7$ & $1.341$ & $95\%$ & \cellcolor[HTML]{EEEEEE}{$15\%$} & $18\%$ \\ 
8 hours & $2.5$ & $4.0$ & $6.4$ & \cellcolor[HTML]{EEEEEE}{$1.301$} & \cellcolor[HTML]{EEEEEE}{$95\%$} & $15\%$ & $20\%$ \\ 
\midrule\addlinespace[2.5pt]
\multicolumn{8}{l}{Time Window} \\ 
\midrule\addlinespace[2.5pt]
< 1 hour & $2.3$ & \cellcolor[HTML]{EEEEEE}{$3.7$} & $5.9$ & \cellcolor[HTML]{EEEEEE}{$1.300$} & $95\%$ & \cellcolor[HTML]{EEEEEE}{$14\%$} & $18\%$ \\ 
< 3 hours & $1.6$ & $3.9$ & $5.5$ & $1.320$ & $95\%$ & $15\%$ & $17\%$ \\ 
No Window & \cellcolor[HTML]{EEEEEE}{$1.3$} & $4.3$ & \cellcolor[HTML]{EEEEEE}{$4.9$} & $1.340$ & \cellcolor[HTML]{EEEEEE}{$95\%$} & $16\%$ & \cellcolor[HTML]{EEEEEE}{$17\%$} \\ 
\midrule\addlinespace[2.5pt]
\multicolumn{8}{l}{Imputations} \\ 
\midrule\addlinespace[2.5pt]
1 & $2.0$ & \cellcolor[HTML]{EEEEEE}{$3.8$} & $5.5$ & \cellcolor[HTML]{EEEEEE}{$1.306$} & $95\%$ & \cellcolor[HTML]{EEEEEE}{$15\%$} & $17\%$ \\ 
3 & \cellcolor[HTML]{EEEEEE}{$1.5$} & $4.1$ & $5.4$ & $1.326$ & $95\%$ & $15\%$ & \cellcolor[HTML]{EEEEEE}{$17\%$} \\ 
5 & $1.6$ & $4.0$ & $5.4$ & $1.325$ & $95\%$ & $15\%$ & $17\%$ \\ 
10 & $1.7$ & $4.0$ & \cellcolor[HTML]{EEEEEE}{$5.4$} & $1.322$ & \cellcolor[HTML]{EEEEEE}{$95\%$} & $15\%$ & $17\%$ \\ 
\bottomrule
\end{longtable}

\begin{landscape}

\begin{figure}

{\centering \includegraphics[width=1\linewidth,height=1\textheight]{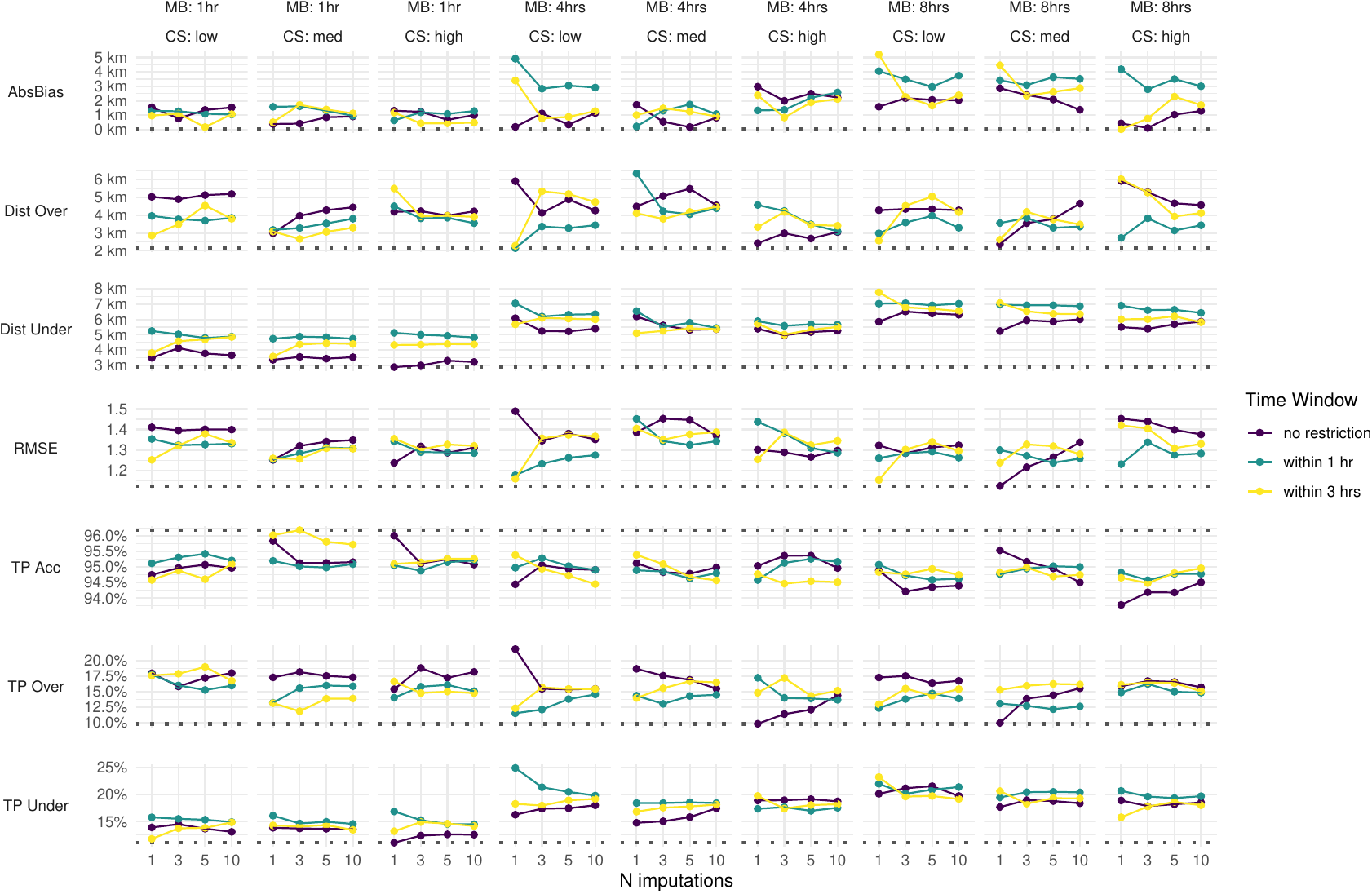} 

}

\caption[This is the caption List of figures]{Visual comparison of DTWBMI parameters across six measures of performance. Match Buffer (MB) and Candidate Specificity (CS) are represented by facets. The Time Window parameter is represented by shape and color. The dotted gray line indicates the best relative performance across all simulation parameter combinations within each measure.}\label{fig:appfig1}
\end{figure}

\end{landscape}

The performance of these metrics is dependent both upon the other
metrics in the simulation model as well as characteristics of the data
and its missingness. Figure \ref{fig:appfig1} shows mean absolute bias,
RMSE and similarity values for the imputation of total distance
traveled. The figure shows the interactions in these metrics across all
parameter values for candidate specificity, match buffer, time window
and number of imputations. The dashed line demonstrates best relative
performance.

We first consider absolute bias (AbsBias) in Figure \ref{fig:appfig1}.
Match buffers (MB) of one hour demonstrate the lowest absolute bias in
meters, with a corresponding increase of absolute bias as the length of
the match buffer increases. A low candidate specificity (CS) has an
increased bias relative to medium and high when the match buffer
increases. Imputing with a match buffer of 8 hours is less biased with a
higher candidate specificity, while imputing with a match buffer of 4
hours is less biased with medium candidate specificity. Differences
between the various time windows emerge across the different match
buffer levels. As the length of the match buffer increases, the 1-hour
time window has reduced performance with respect to bias. The number of
imputations does not have a consistent relationship with the bias.

Next, we consider the distance over- and underestimated across the
various parameters. As is shown in Figure \ref{fig:appfig1}, the general
relationship is that parameters combinations that decrease
overestimation increase underestimation. Conversely, parameter
combinations that decrease underestimation, increase overestimation. As
the match buffer length increases, the underestimation of travel
distance increases. However, the reverse trend is less clear, as a match
buffer of 8 hours does not provide in aggregate a low level of distance
overestimation, but rather ranges between 3 and 6 kilometers on average.
The relationship of candidate specificity level to bias is unclear. An
unrestricted time window provides less underestimation across all
levels, while a 1-hour time window increases underestimation. A 1-hour
time window does not, however, appear to offer the best performance with
respect to overestimation, as this varies across the other parameters.
Lastly, we identify no relationship between the number of imputations
and the underestimation, but a slight increasing relationship between
the number of imputations within the medium candidate specificity
condition.

Travel period accuracy (TP Acc) remains a relatively consistent 95\%
across most condition combinations. When the match buffer is 8 hours, an
unrestricted time window seems to offer worse performance in this
metric. In aggregate, a match buffer of 1 hour offers the best
performance. No clear patterns emerge for candidate specificity or
number of imputations.

Travel period over- and underestimation differ slightly from the
interpretation of the total distance over- and underestimated. The
relationship for match buffer is largely the same, with a modest
increase in rate of travel period underestimation in the 4-hour and
8-hour conditions as compared to the 1-hour condition. Similarly, there
is a reduction in aggregate in travel period overestimation for match
buffers of 4 and 8 hours relative to 1 hour. No clear pattern is evident
for candidate specificity. On average, unrestricted time windows provide
more overestimation and less underestimation, while the most restrictive
time window condition provides less overestimation and more
underestimation of the total number of travel periods. No clear pattern
emerges for number of imputations.

\newpage

\newpage

\hypertarget{sec:appendixb}{%
\section*{Appendix B - Impact of number of own
sets}\label{sec:appendixb}}
\addcontentsline{toc}{section}{Appendix B - Impact of number of own
sets}

Given the assumption that an individual's prior travel behavior often
serves as the most appropriate imputation candidate for their own
missing data, we executed an additional simulation study. This was done
with the aim of evaluating the effects of increasing the availability of
self-referential data sets, or ``own sets''.

We selected the full subset of 16 participants who had a minimum of four
available sets. For each of these individuals, we randomly selected one
set into which missing data was inserted. This missing data was
introduced as a single contiguous block, initiating at a randomly chosen
suitable location within the set. The missingness was introduced at five
different levels, namely one hour, three hours, six hours, eight hours
and twelve hours, thereby establishing five distinct gap lengths within
the missingness condition.

In the own data condition, we created four levels: zero self-sets, one
self-set, two self-sets and three self-sets. Each own data condition
incorporated a base reference set consisting of 48 randomly selected
sets from individuals with fewer than four datasets. This was done to
ensure an ample quantity of imputation candidates. For each condition,
the unused self-sets were replaced with additional reference sets from
individuals having fewer than four data sets, in order to maintain a
consistent number of total available candidates.

We conducted ten simulations wherein DTWBMI-HI, DTWBMI-LO and DTWBI were
applied to the generated data set. These simulations were evaluated
based on the performance criteria delineated in section
\ref{sec:perfcrit} to determine their relative performance on travel
distance imputation.

Table 8 shows aggregate results across the own set condition. Both
DTWBMI-HI and DTWBMI-LO show a reduction in absolute bias corresponding
to an increase in the number of own sets. RMSE and TP Over, TP Under,
and TP Acc. remain mostly stable. The distance overestimated increases
slightly, while the distance underestimated decreases slightly for both
methods. Median bias is excluded from this table, but was 0 across both
methods and all conditions. Overall, it seems that increasing the number
of own sets generally improves the performance of both imputation
methods, with DTWBMI-LO consistently performing better than DTWBMI-HI
based on the presented metrics.

\begin{longtable}{l|rrrrrrr}
\caption*{
{\large Table 8: Method comparison across number of own reference sets}
} \\ 
\toprule
\multicolumn{1}{l}{} & Abs bias & RMSE & TP over & TP under & TP acc. & Dist over & Dist under \\ 
\midrule\addlinespace[2.5pt]
\multicolumn{8}{l}{0} \\ 
\midrule\addlinespace[2.5pt]
DTWBMI-HI & $5.6$ Km & $33.16$ & $14.0\%$ & $17.2\%$ & $94.1\%$ & $4.4$ Km & $10.0$ Km \\ 
DTWBMI-LO & \cellcolor[HTML]{EEEEEE}{$3.9$ Km} & \cellcolor[HTML]{EEEEEE}{$24.88$} & \cellcolor[HTML]{EEEEEE}{$11.4\%$} & \cellcolor[HTML]{EEEEEE}{$12.9\%$} & \cellcolor[HTML]{EEEEEE}{$95.6\%$} & \cellcolor[HTML]{EEEEEE}{$3.1$ Km} & \cellcolor[HTML]{EEEEEE}{$7.0$ Km} \\ 
\midrule\addlinespace[2.5pt]
\multicolumn{8}{l}{1} \\ 
\midrule\addlinespace[2.5pt]
DTWBMI-HI & $4.4$ Km & $34.43$ & $14.6\%$ & $16.4\%$ & $94.1\%$ & $5.2$ Km & $9.6$ Km \\ 
DTWBMI-LO & \cellcolor[HTML]{EEEEEE}{$3.0$ Km} & \cellcolor[HTML]{EEEEEE}{$23.12$} & \cellcolor[HTML]{EEEEEE}{$11.7\%$} & \cellcolor[HTML]{EEEEEE}{$12.1\%$} & \cellcolor[HTML]{EEEEEE}{$95.6\%$} & \cellcolor[HTML]{EEEEEE}{$3.4$ Km} & \cellcolor[HTML]{EEEEEE}{$6.4$ Km} \\ 
\midrule\addlinespace[2.5pt]
\multicolumn{8}{l}{2} \\ 
\midrule\addlinespace[2.5pt]
DTWBMI-HI & $3.8$ Km & $32.69$ & $13.9\%$ & $15.7\%$ & $94.2\%$ & $5.0$ Km & $8.8$ Km \\ 
DTWBMI-LO & \cellcolor[HTML]{EEEEEE}{$2.2$ Km} & \cellcolor[HTML]{EEEEEE}{$24.36$} & \cellcolor[HTML]{EEEEEE}{$12.1\%$} & \cellcolor[HTML]{EEEEEE}{$11.9\%$} & \cellcolor[HTML]{EEEEEE}{$95.6\%$} & \cellcolor[HTML]{EEEEEE}{$3.9$ Km} & \cellcolor[HTML]{EEEEEE}{$6.2$ Km} \\ 
\midrule\addlinespace[2.5pt]
\multicolumn{8}{l}{3} \\ 
\midrule\addlinespace[2.5pt]
DTWBMI-HI & $3.4$ Km & $30.65$ & $13.5\%$ & $16.6\%$ & $94.3\%$ & $5.1$ Km & $8.5$ Km \\ 
DTWBMI-LO & \cellcolor[HTML]{EEEEEE}{$2.1$ Km} & \cellcolor[HTML]{EEEEEE}{$23.69$} & \cellcolor[HTML]{EEEEEE}{$12.1\%$} & \cellcolor[HTML]{EEEEEE}{$12.1\%$} & \cellcolor[HTML]{EEEEEE}{$95.6\%$} & \cellcolor[HTML]{EEEEEE}{$3.8$ Km} & \cellcolor[HTML]{EEEEEE}{$6.0$ Km} \\ 
\bottomrule
\end{longtable}

Table 9 shows the breakout across both gap length condition and number
of self-sets available. As gap length increases, the performance boost
for both DTWBMI methods increases in total bias reduction. Similar
patterns are demonstrated within each level of missing data: Absolute
bias reduces with the increase of the number of own sets, RMSE remains
stable or shows evidence of a slight decrease, and Travel Period metrics
remain largely stable. Overall, DTWBMI-LO outperforms DTWBMI-LO.

\begin{longtable}{l|rrrrrrrr}
\caption*{
{\large Table 9: Method comparison across gap length and own reference sets}
} \\ 
\toprule
\multicolumn{1}{l}{} & Own sets & Abs bias & RMSE & TP over & TP under & TP acc. & Dist over & Dist under \\ 
\midrule\addlinespace[2.5pt]
\multicolumn{9}{l}{1h} \\ 
\midrule\addlinespace[2.5pt]
DTWBMI-HI & 0 & $1.7$ Km & $11.9$ & $3.7\%$ & $5.7\%$ & $94.6\%$ & $0.4$ Km & $2.1$ Km \\ 
DTWBMI-HI & 1 & $1.7$ Km & $11.3$ & \cellcolor[HTML]{EEEEEE}{$2.7\%$} & $4.9\%$ & $95.7\%$ & $0.2$ Km & $2.0$ Km \\ 
DTWBMI-HI & 2 & $2.0$ Km & $11.8$ & $3.8\%$ & $5.6\%$ & $94.9\%$ & $0.3$ Km & $2.3$ Km \\ 
DTWBMI-HI & 3 & $2.0$ Km & $12.0$ & \cellcolor[HTML]{EEEEEE}{$2.5\%$} & $5.8\%$ & $95.5\%$ & $0.2$ Km & $2.2$ Km \\ 
DTWBMI-LO & 0 & \cellcolor[HTML]{EEEEEE}{$1.3$ Km} & \cellcolor[HTML]{EEEEEE}{$9.9$} & \cellcolor[HTML]{EEEEEE}{$2.1\%$} & \cellcolor[HTML]{EEEEEE}{$2.7\%$} & \cellcolor[HTML]{EEEEEE}{$97.3\%$} & \cellcolor[HTML]{EEEEEE}{$0.1$ Km} & \cellcolor[HTML]{EEEEEE}{$1.4$ Km} \\ 
DTWBMI-LO & 1 & \cellcolor[HTML]{EEEEEE}{$1.3$ Km} & \cellcolor[HTML]{EEEEEE}{$9.4$} & $2.7\%$ & \cellcolor[HTML]{EEEEEE}{$2.3\%$} & \cellcolor[HTML]{EEEEEE}{$97.3\%$} & \cellcolor[HTML]{EEEEEE}{$0.1$ Km} & \cellcolor[HTML]{EEEEEE}{$1.4$ Km} \\ 
DTWBMI-LO & 2 & \cellcolor[HTML]{EEEEEE}{$0.8$ Km} & \cellcolor[HTML]{EEEEEE}{$7.2$} & \cellcolor[HTML]{EEEEEE}{$2.4\%$} & \cellcolor[HTML]{EEEEEE}{$1.8\%$} & \cellcolor[HTML]{EEEEEE}{$97.8\%$} & \cellcolor[HTML]{EEEEEE}{$0.2$ Km} & \cellcolor[HTML]{EEEEEE}{$1.0$ Km} \\ 
DTWBMI-LO & 3 & \cellcolor[HTML]{EEEEEE}{$0.8$ Km} & \cellcolor[HTML]{EEEEEE}{$7.8$} & $2.9\%$ & \cellcolor[HTML]{EEEEEE}{$2.1\%$} & \cellcolor[HTML]{EEEEEE}{$97.4\%$} & \cellcolor[HTML]{EEEEEE}{$0.2$ Km} & \cellcolor[HTML]{EEEEEE}{$1.0$ Km} \\ 
\midrule\addlinespace[2.5pt]
\multicolumn{9}{l}{3h} \\ 
\midrule\addlinespace[2.5pt]
DTWBMI-HI & 0 & $2.8$ Km & $19.4$ & $7.1\%$ & $12.5\%$ & $95.3\%$ & $1.1$ Km & $4.0$ Km \\ 
DTWBMI-HI & 1 & $2.9$ Km & $18.5$ & $8.5\%$ & $13.9\%$ & $94.6\%$ & \cellcolor[HTML]{EEEEEE}{$1.2$ Km} & $4.1$ Km \\ 
DTWBMI-HI & 2 & $2.7$ Km & $17.0$ & \cellcolor[HTML]{EEEEEE}{$6.0\%$} & $12.4\%$ & $95.6\%$ & \cellcolor[HTML]{EEEEEE}{$1.0$ Km} & $3.7$ Km \\ 
DTWBMI-HI & 3 & $2.4$ Km & $19.0$ & \cellcolor[HTML]{EEEEEE}{$7.3\%$} & $12.4\%$ & $95.1\%$ & $1.5$ Km & $3.8$ Km \\ 
DTWBMI-LO & 0 & \cellcolor[HTML]{EEEEEE}{$1.5$ Km} & \cellcolor[HTML]{EEEEEE}{$12.5$} & \cellcolor[HTML]{EEEEEE}{$6.8\%$} & \cellcolor[HTML]{EEEEEE}{$8.4\%$} & \cellcolor[HTML]{EEEEEE}{$95.9\%$} & \cellcolor[HTML]{EEEEEE}{$1.0$ Km} & \cellcolor[HTML]{EEEEEE}{$2.4$ Km} \\ 
DTWBMI-LO & 1 & \cellcolor[HTML]{EEEEEE}{$0.6$ Km} & \cellcolor[HTML]{EEEEEE}{$9.7$} & \cellcolor[HTML]{EEEEEE}{$6.5\%$} & \cellcolor[HTML]{EEEEEE}{$8.4\%$} & \cellcolor[HTML]{EEEEEE}{$96.0\%$} & $1.3$ Km & \cellcolor[HTML]{EEEEEE}{$1.9$ Km} \\ 
DTWBMI-LO & 2 & \cellcolor[HTML]{EEEEEE}{$0.5$ Km} & \cellcolor[HTML]{EEEEEE}{$11.0$} & $8.4\%$ & \cellcolor[HTML]{EEEEEE}{$8.4\%$} & \cellcolor[HTML]{EEEEEE}{$95.8\%$} & $1.4$ Km & \cellcolor[HTML]{EEEEEE}{$1.9$ Km} \\ 
DTWBMI-LO & 3 & \cellcolor[HTML]{EEEEEE}{$0.6$ Km} & \cellcolor[HTML]{EEEEEE}{$10.9$} & $7.9\%$ & \cellcolor[HTML]{EEEEEE}{$7.8\%$} & \cellcolor[HTML]{EEEEEE}{$95.8\%$} & \cellcolor[HTML]{EEEEEE}{$1.3$ Km} & \cellcolor[HTML]{EEEEEE}{$1.9$ Km} \\ 
\midrule\addlinespace[2.5pt]
\multicolumn{9}{l}{6h} \\ 
\midrule\addlinespace[2.5pt]
DTWBMI-HI & 0 & $6.1$ Km & $27.3$ & $13.4\%$ & $18.4\%$ & $94.4\%$ & $2.6$ Km & $8.6$ Km \\ 
DTWBMI-HI & 1 & $6.0$ Km & $26.6$ & $14.5\%$ & $19.1\%$ & $94.2\%$ & $2.6$ Km & $8.6$ Km \\ 
DTWBMI-HI & 2 & $4.0$ Km & $29.4$ & $17.5\%$ & $18.1\%$ & $93.5\%$ & $4.3$ Km & $8.3$ Km \\ 
DTWBMI-HI & 3 & $4.0$ Km & $24.0$ & $13.6\%$ & $18.2\%$ & $94.1\%$ & $3.3$ Km & $7.2$ Km \\ 
DTWBMI-LO & 0 & \cellcolor[HTML]{EEEEEE}{$3.0$ Km} & \cellcolor[HTML]{EEEEEE}{$17.2$} & \cellcolor[HTML]{EEEEEE}{$12.0\%$} & \cellcolor[HTML]{EEEEEE}{$14.7\%$} & \cellcolor[HTML]{EEEEEE}{$95.2\%$} & \cellcolor[HTML]{EEEEEE}{$2.3$ Km} & \cellcolor[HTML]{EEEEEE}{$5.3$ Km} \\ 
DTWBMI-LO & 1 & \cellcolor[HTML]{EEEEEE}{$1.9$ Km} & \cellcolor[HTML]{EEEEEE}{$16.9$} & \cellcolor[HTML]{EEEEEE}{$10.8\%$} & \cellcolor[HTML]{EEEEEE}{$12.7\%$} & \cellcolor[HTML]{EEEEEE}{$95.5\%$} & \cellcolor[HTML]{EEEEEE}{$2.5$ Km} & \cellcolor[HTML]{EEEEEE}{$4.4$ Km} \\ 
DTWBMI-LO & 2 & \cellcolor[HTML]{EEEEEE}{$1.1$ Km} & \cellcolor[HTML]{EEEEEE}{$17.2$} & \cellcolor[HTML]{EEEEEE}{$12.8\%$} & \cellcolor[HTML]{EEEEEE}{$12.1\%$} & \cellcolor[HTML]{EEEEEE}{$95.6\%$} & \cellcolor[HTML]{EEEEEE}{$3.0$ Km} & \cellcolor[HTML]{EEEEEE}{$4.1$ Km} \\ 
DTWBMI-LO & 3 & \cellcolor[HTML]{EEEEEE}{$1.4$ Km} & \cellcolor[HTML]{EEEEEE}{$17.1$} & \cellcolor[HTML]{EEEEEE}{$11.6\%$} & \cellcolor[HTML]{EEEEEE}{$12.8\%$} & \cellcolor[HTML]{EEEEEE}{$95.5\%$} & \cellcolor[HTML]{EEEEEE}{$2.8$ Km} & \cellcolor[HTML]{EEEEEE}{$4.2$ Km} \\ 
\midrule\addlinespace[2.5pt]
\multicolumn{9}{l}{8h} \\ 
\midrule\addlinespace[2.5pt]
DTWBMI-HI & 0 & $3.2$ Km & $40.6$ & $22.4\%$ & $21.5\%$ & $93.9\%$ & $7.9$ Km & $11.1$ Km \\ 
DTWBMI-HI & 1 & \cellcolor[HTML]{EEEEEE}{$0.2$ Km} & $50.6$ & $22.1\%$ & $19.2\%$ & $93.6\%$ & $10.7$ Km & $10.9$ Km \\ 
DTWBMI-HI & 2 & $1.3$ Km & $44.9$ & $20.0\%$ & $17.7\%$ & $94.0\%$ & $9.1$ Km & $10.4$ Km \\ 
DTWBMI-HI & 3 & $2.0$ Km & $38.7$ & $21.6\%$ & $21.2\%$ & $94.1\%$ & $8.2$ Km & $10.2$ Km \\ 
DTWBMI-LO & 0 & \cellcolor[HTML]{EEEEEE}{$1.2$ Km} & \cellcolor[HTML]{EEEEEE}{$26.7$} & \cellcolor[HTML]{EEEEEE}{$17.7\%$} & \cellcolor[HTML]{EEEEEE}{$15.9\%$} & \cellcolor[HTML]{EEEEEE}{$95.4\%$} & \cellcolor[HTML]{EEEEEE}{$6.0$ Km} & \cellcolor[HTML]{EEEEEE}{$7.2$ Km} \\ 
DTWBMI-LO & 1 & $0.9$ Km & \cellcolor[HTML]{EEEEEE}{$25.2$} & \cellcolor[HTML]{EEEEEE}{$19.5\%$} & \cellcolor[HTML]{EEEEEE}{$15.3\%$} & \cellcolor[HTML]{EEEEEE}{$95.0\%$} & \cellcolor[HTML]{EEEEEE}{$6.0$ Km} & \cellcolor[HTML]{EEEEEE}{$6.9$ Km} \\ 
DTWBMI-LO & 2 & \cellcolor[HTML]{EEEEEE}{$0.1$ Km} & \cellcolor[HTML]{EEEEEE}{$27.6$} & \cellcolor[HTML]{EEEEEE}{$17.5\%$} & \cellcolor[HTML]{EEEEEE}{$15.9\%$} & \cellcolor[HTML]{EEEEEE}{$95.1\%$} & \cellcolor[HTML]{EEEEEE}{$6.3$ Km} & \cellcolor[HTML]{EEEEEE}{$6.4$ Km} \\ 
DTWBMI-LO & 3 & \cellcolor[HTML]{EEEEEE}{$0.6$ Km} & \cellcolor[HTML]{EEEEEE}{$26.8$} & \cellcolor[HTML]{EEEEEE}{$18.2\%$} & \cellcolor[HTML]{EEEEEE}{$16.9\%$} & \cellcolor[HTML]{EEEEEE}{$95.2\%$} & \cellcolor[HTML]{EEEEEE}{$5.8$ Km} & \cellcolor[HTML]{EEEEEE}{$6.3$ Km} \\ 
\midrule\addlinespace[2.5pt]
\multicolumn{9}{l}{12h} \\ 
\midrule\addlinespace[2.5pt]
DTWBMI-HI & 0 & $14.2$ Km & $66.7$ & $23.5\%$ & $27.7\%$ & $92.5\%$ & $9.8$ Km & $24.0$ Km \\ 
DTWBMI-HI & 1 & $11.3$ Km & $65.2$ & $25.1\%$ & $24.8\%$ & $92.4\%$ & $11.1$ Km & $22.4$ Km \\ 
DTWBMI-HI & 2 & $9.1$ Km & $60.3$ & $22.2\%$ & $24.7\%$ & $92.8\%$ & $10.5$ Km & $19.6$ Km \\ 
DTWBMI-HI & 3 & \cellcolor[HTML]{EEEEEE}{$6.5$ Km} & $59.6$ & $22.3\%$ & $25.5\%$ & $92.8\%$ & $12.3$ Km & $18.8$ Km \\ 
DTWBMI-LO & 0 & \cellcolor[HTML]{EEEEEE}{$12.7$ Km} & \cellcolor[HTML]{EEEEEE}{$58.1$} & \cellcolor[HTML]{EEEEEE}{$18.5\%$} & \cellcolor[HTML]{EEEEEE}{$22.8\%$} & \cellcolor[HTML]{EEEEEE}{$94.1\%$} & \cellcolor[HTML]{EEEEEE}{$6.2$ Km} & \cellcolor[HTML]{EEEEEE}{$18.8$ Km} \\ 
DTWBMI-LO & 1 & \cellcolor[HTML]{EEEEEE}{$10.3$ Km} & \cellcolor[HTML]{EEEEEE}{$54.4$} & \cellcolor[HTML]{EEEEEE}{$18.8\%$} & \cellcolor[HTML]{EEEEEE}{$21.8\%$} & \cellcolor[HTML]{EEEEEE}{$94.0\%$} & \cellcolor[HTML]{EEEEEE}{$7.1$ Km} & \cellcolor[HTML]{EEEEEE}{$17.4$ Km} \\ 
DTWBMI-LO & 2 & \cellcolor[HTML]{EEEEEE}{$8.5$ Km} & \cellcolor[HTML]{EEEEEE}{$58.9$} & \cellcolor[HTML]{EEEEEE}{$19.2\%$} & \cellcolor[HTML]{EEEEEE}{$21.2\%$} & \cellcolor[HTML]{EEEEEE}{$93.9\%$} & \cellcolor[HTML]{EEEEEE}{$8.9$ Km} & \cellcolor[HTML]{EEEEEE}{$17.4$ Km} \\ 
DTWBMI-LO & 3 & $7.4$ Km & \cellcolor[HTML]{EEEEEE}{$55.8$} & \cellcolor[HTML]{EEEEEE}{$19.9\%$} & \cellcolor[HTML]{EEEEEE}{$20.7\%$} & \cellcolor[HTML]{EEEEEE}{$93.9\%$} & \cellcolor[HTML]{EEEEEE}{$9.1$ Km} & \cellcolor[HTML]{EEEEEE}{$16.5$ Km} \\ 
\bottomrule
\end{longtable}

Although we expected DTWBMI-HI to outperform DTWBMI-LO in situations
where it was advantageous to match with a greater specificity to the
shape of the behavior, this simulation study does not provide evidence
indicating that a small increase in the available historical data for a
person would be sufficient to prefer the use of the high-information
method to the low-information method.

\end{document}